\documentclass[
aps,prd,
10pt,
superscriptaddress,
nofootinbib,
amsmath,amssymb
]{revtex4-2}
\usepackage{upgreek, physics, stix, dsfont, bm} 

\usepackage{bbold} 
\usepackage{comment} 

\usepackage{graphicx, xcolor}
\definecolor{maroon}{HTML}{800020}
\definecolor{darkcyan}{HTML}{008060}
\usepackage[colorlinks]{hyperref} 
\hypersetup{colorlinks=true, allcolors=maroon}
\usepackage[capitalise]{cleveref} 
\crefname{section}{Sec.}{Secs.} 

\newcommand{\ii}{\ensuremath{\mathrm{i}}}
\newcommand{\ee}{\ensuremath{\mathrm{e}}}

\DeclareMathOperator{\sgn}{sgn}

\begin{document} 


\title{Emergent metrics from wavelet-transformed quantum field theory}

\newcommand{\affmq}{School of Mathematical and Physical Sciences, Macquarie University, Sydney, NSW 2109, Australia}
\newcommand{\affequs}{ARC Centre of Excellence in Engineered Quantum Systems, Macquarie University, Sydney, NSW 2109, Australia}
\newcommand{\affsqa}{Sydney Quantum Academy, Sydney, NSW 2000, Australia}
\newcommand{\affrmit}{Centre for Quantum Computation and Communication Technology, School of Science, RMIT University, Melbourne, VIC 3000, Australia}
\newcommand{\affuwat}{Department of Applied Mathematics and Department of Physics, University of Waterloo, Waterloo N2L 3G1, Canada}
\newcommand{\affpitp}{Perimeter Institute for Theoretical Physics, Waterloo N2L 2Y5, Canada}

\date{December 2025}

\author{Šimon~Vedl}
\email[Please direct correspondence to: ]{simon.vedl@hdr.mq.edu.au}
\author{Daniel~J.~George}
\affiliation{\affmq}
\affiliation{\affequs}
\affiliation{\affsqa}

\author{Fil~Simovic}
\affiliation{\affmq}
\affiliation{\affuwat}

\author{Dominic~G.~Lewis}
\author{Nicholas Funai}
\affiliation{\affrmit}

\author{Achim~Kempf}
\affiliation{\affuwat}
\affiliation{\affpitp}

\author{Nicolas~C.~Menicucci}
\affiliation{\affrmit}

\author{Gavin~K.~Brennen}
\affiliation{\affmq}
\affiliation{\affequs}

\begin{abstract}
We introduce a method of reverse holography by which a bulk metric is shown to arise from locally computable multiscale correlations of a boundary quantum field theory (QFT). 
The metric is obtained from the Petz-Rényi mutual information defined with input correlations computed from the continuous wavelet transform.
The method is applicable to a variety of boundary QFTs that need not be conformal field theories (CFTs).
For thermal free scalar and Dirac field theories the resulting bulk metric is that of (asymmetrically) warped anti-de Sitter (AdS) space.
For massless, ground state CFTs the geometry simply reduces to AdS space.
We show that certain parameters of the geometry can be tuned by changing the shape of the wavelet function.
\end{abstract}

\maketitle


\section{Introduction}
\label{sec:introduction}

The holographic principle has been a topic of significant interest for several decades now, emerging from studies of black hole entropy in the 1990s \cite{hooft1993,susskind1995} that have developed into numerous concrete realizations today. The central idea of holography is that all of the information contained in a system can be encoded on a (suitably chosen) lower dimensional boundary of that system. The most famous example of such a duality is the anti-de Sitter space/conformal field theory (AdS/CFT) correspondence, which conjectures a one-to-one correspondence between states and observables in string theory on $(1 + d + 1)$-dimensional asymptotically AdS spaces and states in certain conformal field theories (CFTs) on $(1 + d)$-dimensional spaces \cite{Maldacena1999, Witten:1998qj}. CFTs themselves are of broad interest as they can be used to describe quantum critical points, systems in a critical phase such as a massless phase of a scalar field, defects in materials, and more. In some cases, the mapping of CFT data to geometric features of the bulk description enable dramatic simplifications to otherwise difficult computations \cite{ryu2006c}. In other cases, new insights can be gained about systems at strong coupling by taking advantage of the strong-weak coupling duality of AdS/CFT \cite{buchel2005b, kovtun2005b}.

The AdS/CFT correspondence also finds application in quantum information science. For example, to correct errors in quantum computation, a class of holographic quantum error correcting codes has been proposed \cite{pastawski2015}. The performance of these codes can be analysed using tools from the AdS/CFT correspondence, for example, the ability to recover from erasure errors can be quantified using the entanglement wedge. Another example, closely related to the present work, is the multiscale entanglement renormalisation ansatz (MERA) \cite{Vidal:2007}. This is a tensor network technique that can be used to simulate highly entangled states including vacuum states of CFTs.  The tensor network, which is an approximate representation of a discretized boundary field, itself resembles a tessellation of a time slice of AdS spacetime, suggesting some form of bulk-boundary correspondence \cite{Swingle:2012}. And more generally, $n$-ary tree graphs, such as those representing tensor networks, can also be related to hyperbolic geometry \cite{altaisky2022}. MERA networks can be ``lifted" by interpreting the network bonds and tensors as physical degrees of freedom in the bulk, revealing several features common to other approaches to holography: holographic screens, mappings between bulk and boundary operators, and bulk gauging of global on-site symmetries on the boundary \cite{Singh:2018, McMahon2020}. However, while MERA tensor networks explicitly break the boundary spatial symmetry in the bulk, continuous MERA (cMERA) \cite{Haegeman:2013} provides a way to restore this by working with a generator of coarse-graining, continuously parameterised by scale.
This tool was used effectively in one of the first attempts to explicitly obtain the metric of AdS from purely quantum field theoretic data in \textcite{Nozaki2012}.

From the perspective of holography, a shortcoming of tensor network approaches is that the tensors must be computed via a minimization procedure and typically represent only an equal-time slice of the boundary. The wavelet transform offers an alternative mathematical tool for constructing multiscale representations and scale dependent features of QFT in space-time. Wavelets provide a regularised description of QFTs with an additional scale label \cite{altaisky2012,altaisky2013}. Wavelets are the primary tool appearing in the exact holographic mapping \cite{qi2013,LeeQi:2016,SinghBrennen:2016,lee2017}, in which a discrete basis of wavelets is used to construct a multiscale representation of a CFT. In this example, the \emph{long} distance behaviour of mutual information between quantum states with support on subsystems defined on discrete points in space and scale and continuous points in time appears to follow the geodesic distance of AdS with a radius of curvature that depends on the choice of wavelet basis \cite{qi2013, SinghBrennen:2016,lee2017,LeeQi:2016}. Additionally, the wavelet description provides a means to compress a many-body quantum state thus simplifying otherwise difficult-to-calculate quantities like holographic entanglement of purification \cite{george2022}. These observations suggest that the wavelet description could capture some aspects of the AdS/CFT correspondence \cite{lee2017}. However, these results arise from lattice descriptions which only qualitatively suggest existence of an AdS-like manifold. \footnote{It is also interesting to note that in these works \cite{qi2013,LeeQi:2016,SinghBrennen:2016,lee2017} a signature change has been observed in the metric, where a Riemannian metric is obtained instead of a Lorentzian one.} A more substantive connection between AdS/CFT and holographic features appearing in wavelet decompositions of QFTs could be uncovered if one could construct a metric tensor for an asymptotically AdS manifold with the original spacetime coordinates plus an additional scale coordinate. This would enable the direct computation of various geometric quantities in the bulk, facilitating a more precise understanding of the dictionary between wavelet modes and bulk geometry. We are thus naturally led to consider the continuous wavelet transform, where both the position and scale of the wavelet are described by continuous variables.

To construct explicit metric tensors from field theory data, we take inspiration from recent proposals by \textcite{saravani2016} and \textcite{kempf2021} who suggest a method for reconstructing the metric purely from the correlations present in a QFT. \textcite{kempf2021} further adopts the view that quantum field correlations may be fundamental and that manifolds and metrics exist only insofar as they are derived from the $n$-point functions of the quantum field. More concretely, \textcite{perche2022} provided a scheme for reconstructing a spacetime manifold from measurements performed by local detectors. This philosophy suits our task, because the wavelet description of a QFT can be viewed as a multiscale decomposition with respect to detectors at different resolutions (scale), and there is no implicit notion of manifold or metric.

In this work we begin by adopting the schema of \textcite{kempf2021}, extending it to encompass fermionic QFT and computing the metric directly from multiscale field theory correlations. We show that, naively, the different scaling dimensions of the operators present in the theory prevent one from constructing a sensible metric directly from the two-point correlation function. We remedy this by adopting a locally basis-independent function in the form of the Petz-Rényi mutual information (PRMI), an entropic analogue of the two-point function. This choice is motivated by the observation that for a CFT, the PRMI can be computed from quantum correlations in the continuum setting and is not ultraviolet (UV) divergent \cite{kudler-flam2023}. We then demonstrate that this technique can be used for non-conformal theories as well, by computing the bulk metric `dual' to vacuum and thermal states of both massive and massless free Dirac fields in $1+1$ dimensions and free scalar fields in $1+d$ dimensions from their respective PRMIs. We find that the bulk metric is asymptotically AdS with a UV cutoff scale associated with the finest resolution of the detectors, and provide a wavelet basis constructed from affine group coherent states where the radius of curvature is continuously tunable. Finally, we demonstrate that the resulting metrics, which resemble well-known domain wall solutions in AdS/CFT, cannot be explicitly realised as solutions to Einstein-scalar field theory. We conclude with comments about the interpretation of the bulk-boundary correspondence and the holographic nature of wavelet decompositions.


\section{Background}
\label{sec:background}

A natural first approach to constructing a metric with an additional scale dimension is to apply the metric reconstruction technique of \textcite{saravani2016} directly to wavelet-transformed correlation functions. We will demonstrate that this na\"{i}ve approach encounters fundamental obstructions that reveal important constraints on metric reconstruction from multiscale quantum correlations.

Consider the prescription for extracting the $(1+d)$-dimensional metric $g_{\mu\nu}(x)$ (for $d>1$) from the two-point correlation function of a single scalar field $\hat\phi(x)$ living on $g$. The metric is computed from
\begin{equation}
    g_{\mu\nu}(x)
    = -\frac12 \left( \frac{ \Gamma(\frac{d-1}{2}) }{ 4\pi^{\frac{d+1}{2}} } \right)^\frac{2}{d-1}
    \lim_{y \to x} \pdv{x^\mu} \pdv{y^\nu} \left( \expval{\hat{\phi}(x) \hat{\phi}(y)}^\frac{2}{1-d} \right)\ ,
    \label{eq:metric_from_scalar_field_correlations}
\end{equation}
where $x^\mu$, $y^\nu$ are co\"ordinates of the spacetime points $x$ and $y$. As we show in \cref{app:metric_from_dirac}, this can be generalized to a Dirac field $\hat{\bar{\Psi}}(x)$ as well, giving 
\begin{equation}
    g_{\mu\nu}(x) 
    = -\frac{1}{2} \left( \frac{ \Gamma(\frac{d+1}{2}) }{ 2\pi^{\frac{d+1}{2}} } \right)^{\frac{2}{d}}
    \lim_{y\to x} \pdv{x^\mu} \pdv{y^\nu}
    \left[ \frac{\Tr\left(\expval{\hat{\Psi}(x) \hat{\bar{\Psi}}(y)}^2\right) }{\Tr\left(\mathbb{1}_N\right)} \right]^{-\frac{1}{d}},
    \label{eq:metric_from_dirac_field_correlations}
\end{equation}
where $N=2^{\lfloor(1+d)/2\rfloor}$ is the dimension of the corresponding $\gamma$ matrices.
\Cref{eq:metric_from_dirac_field_correlations} usefully also holds for $d=1$.
Note that the formulas \cref{eq:metric_from_scalar_field_correlations} and \cref{eq:metric_from_dirac_field_correlations} assume an arbitrarily high resolution of the field operators.

\begin{figure}[htbp]
    \centering
    \includegraphics[width=.7\columnwidth]{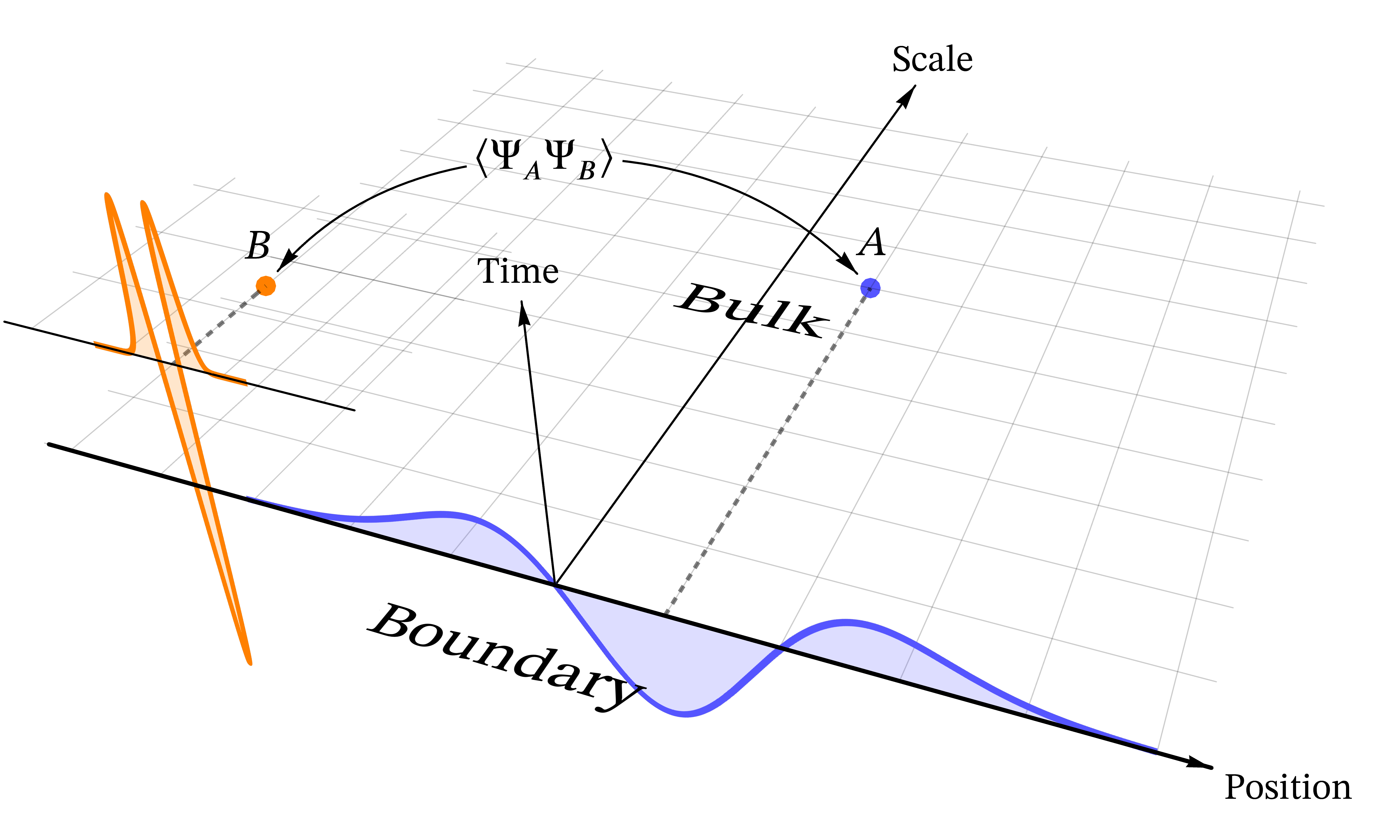}
    \caption{An illustration of the boundary measurements used to infer the bulk metric. Two boundary field operators $\hat{\Psi}_A$ and $\hat{\Psi}_B$ are supported on wavelet modes (indicated by the orange and blue curves) with collective space, time, and scale co\"ordinates $A$ and $B$. The finite-resolution correlation measurement is used as input to the PRMI to compute the metric at $A$ after taking derivatives at the coincidence limit $B \rightarrow A$.}
    \label{fig:scale_distance}
\end{figure}

We will use these expressions to derive the metric of the $(1+2)$-dimensional geometry that emerges from the application of the continuous wavelet transform to a $(1+1)$-dimensional conformal field theory. The use of wavelets allows for UV-friendly descriptions of QFT in a pseudo-local basis at the cost of introducing an additional scale dimension. We consider two cases---a single free massless scalar field and a single free massless Dirac field, both in flat spacetime.

The continuous wavelet transform on $L^2(\mathbb{R})$ is defined as
\begin{equation}
    \phi(x,a) 
    = \braket{w_{a,x}}{\phi} 
    = \int_{-\infty}^{+\infty}w_{a,x}(x')^*\phi(x')\dd{x'}\ ,
\end{equation}
where $w_{a,x}(x')=a^{-1/2}w(\frac{x'-x}{a})$, $w$ is the wavelet function, and $a>0$ is the wavelet scale. 
This can be extended to higher dimensions as
\begin{equation}
    \phi(\bm{x},a,\theta) 
    = \braket{w_{a,\theta,\bm{x}}}{\phi} 
    = \int_{\mathbb{R}^d}w_{a,\theta,\bm{x}}(\bm{x}')^*\phi(\bm{x}')\dd[d]{\bm{x}'} 
    = \int_{\mathbb{R}^d}\ee^{-\ii\bm{p}\cdot\bm{x}}\Tilde{w}_{a,\theta}(\bm{p})^*\Tilde{\phi}(\bm{p})\frac{\dd[d]{\bm{p}}}{(2\pi)^d}\ ,
\end{equation}
where $\theta \in SO(d)$ additionally describes the orientation of the wavelet function (for further details and conditions on the invertibility of the transform see \cref{app:wavelet_transform}).
The representation is faithful when the scale dimension is permitted to run from $0$ (UV limit) to $\infty$ (IR limit).

For simplicity, here we will restrict ourselves to isotropic wavelet functions that are independent of $\theta$, which will be applied to the spatial dimensions only. This ensures that the coincidences for the scalar field (see \cref{app:wavelet_correlators_bosonic}) are bound by the uncertainty principle independent of the dispersion relation:
\begin{equation}
    \expval*{\hat{\phi}(x,a,\theta)^2}\expval*{\hat{\pi}(x,a,\theta)^2}
    = \frac14\norm{\omega_{\bm{p}/a}^{-1/2}\Tilde{w}}^2\norm{\omega_{\bm{p}/a}^{1/2}\Tilde{w}}^2 
    \geq \frac14
\end{equation}
The above implies that every point in the wavelet picture has a physical covariance matrix.
If one were to extend the wavelet transform to also include time (as proposed by \textcite{gorodnitskiy2011,altaisky2012}) the product $\expval*{\hat{\phi}(x,a,\theta)^2}\expval*{\hat{\pi}(x,a,\theta)^2}$ would become proportional to $a^2$, which implies a scale limit beyond which the points in the wavelet picture no longer have a physical covariance matrix.

Substituting the wavelet-transformed correlators into their respective formulas \cref{eq:metric_from_scalar_field_correlations} and \cref{eq:metric_from_dirac_field_correlations} yields metric tensors of the following forms,
\begin{equation}
    \begin{array}{rl}
        \dd{s}^2_{\text{b}}&= \frac{1}{a^4}\left(-\zeta_{tt}\dd{t}^2+\zeta_{xx}\dd{x}^2+\zeta_{aa}\dd{a}^2+\ii \zeta_{ta}(\dd{t}\dd{a}-\dd{a}\dd{t})\right)\ ,\\    
        \dd{s}^2_{\text{f}} &= \frac{1}{a^2}\left(-\xi_{tt}\dd{t}^2+\xi_{xx}\dd{x}^2+\xi_{aa}\dd{a}^2+\ii \xi_{ta}(\dd{t}\dd{a}-\dd{a}\dd{t})\right)\ ,
    \end{array}
\end{equation}
where $\zeta_{\mu\nu}$ and $\xi_{\mu\nu}$ are positive real-valued constants that can be expressed in terms of certain wavelet-dependent coefficients.

Several features of the resulting metrics indicate that the scale dependence of wavelet transformed correlation functions is not straightforwardly encoded in a geometric way. Firstly, the metrics are not diagonal, having purely imaginary off-diagonal components. Secondly, the bosonic field and fermionic fields give distinctly different geometries for the (1+2)-dimensional spacetime, particularly with respect to the scale co\"ordinate $a$, despite considering the same background metric for both fields. Computing the Ricci scalar for the two metric tensors, we obtain a constant negative Ricci scalar for the fermionic theory (consistent with AdS spacetime) and a non-constant Ricci scalar for the bosonic theory. Thus, the direct application of Eqs. \eqref{eq:metric_from_scalar_field_correlations} and \eqref{eq:metric_from_dirac_field_correlations} to wavelet correlators fails to produce consistent geometric structures. We conjecture that the field correlators do not account for the whole operator content of the theory, e.g., we are omitting the correlations of the conjugate field $\hat{\pi}$ which scale differently compared to the scalar field $\hat{\phi}$ itself. What is needed is an operator-basis-independent quantity to compute the metric, which we consider in the following section.

%

\section{Petz-Rényi Mutual Information}
\label{sec:prmi}

A natural choice of quantity that is invariant under a change of basis of the field operators is the Petz-Rényi mutual information (PRMI), denoted $I_\alpha$.
The PRMI is a generalisation of the von Neumann mutual information, and between two subregions $A$ and $B$ of a quantum state, can be expressed as the Petz-Rényi relative entropy (PRRE) of the joint state of the subsystem $\hat{\rho}_{AB}$ given their product state $\hat{\rho}_A \otimes \hat{\rho}_B$:
\begin{equation}
    I_\alpha(A;B) 
    = D_\alpha(\hat{\rho}_{AB} \parallel \hat{\rho}_A \otimes \hat{\rho}_B)
\end{equation}
Similarly, the Petz-Rényi relative entropy is a generalisation of the quantum relative entropy and is defined as
\begin{equation}
    D_\alpha(\hat{\rho} \parallel \hat{\sigma})
    = \frac{1}{\alpha-1} \log \left[ \Tr(\hat{\rho}^\alpha \hat{\sigma}^{1-\alpha}) \right]\ .
\end{equation}
Both the PRMI and PRRE approach the von Neumann mutual information and quantum relative entropy, respectively, in the limit $\alpha \to 1$.
For $\alpha \in [0,2]$, PRRE is useful as a measure of distinguishability between quantum states because it is monotonic under quantum channels \cite{PETZ198657}.
This means that applying a completely positive, trace-preserving quantum channel $\mathcal{E}$ to both states does not make them more distinguishable, that is, $D_\alpha(\hat{\rho} \parallel \hat{\sigma})\geq D_\alpha(\mathcal{E}(\hat{\rho}) \parallel \mathcal{E}(\hat{\sigma}))$.

In our case, we consider $A$ and $B$ to be point-like subsystems on the same time slice in the wavelet picture with co\"ordinates $A=(x,a)$ and $B=(y,b)$.
To avoid direct use of ill-defined density operators in field theory we will rely on the quadratic nature of free field theories, i.e that the vacuum and thermal states are Gaussian states and fully describable by their covariance matrices.
This allows us to use existing results by \textcite{casini2018}.
We will consider the PRMI with parameter $\alpha=2$ because it remains well-defined even when the covariance matrices are non-physical, something which inevitably occurs in the coincidence limit when $A$ and $B$ are brought infinitesimally close together due to the non-orthogonality of the wavelet modes. This choice also has the advantage of being more tractable computationally. The expressions for the PRMI in terms of two-point correlation functions is given in \cref{app:prmi}.
The result can further be extended in the time direction by replacing the equal-time correlation functions with general correlation functions with non-equal times.
When the times are not equal, interpreting this function as a mutual information proves challenging due to the fact that there is no clear interpretation for the joined state $\hat{\rho}_{AB}$ across two time slices.

%

\section{PRMI-derived metric for free Dirac fermion}
\label{sec:prmi_metric_fermionic}
We first consider a free Dirac field in a thermal state in $(1+1)$ dimensions (which is a Gaussian state).The Petz-Rényi mutual information with $\alpha = 2$ between subsystems $A$$:(t, x, a)$ and $B$$:(t', y, b)$ is (see \cref{app:prmi_fermionic} for details):
\begin{equation}
    I_2((t, x, a);(t', y, b))
    = \log(\frac{\det(C_\text{j}^2-2C_\text{j}C_\text{p}+C_\text{p})}{\det(C_\text{p}(1-C_\text{p}))})
\end{equation}
where
\begin{equation}
    C_\text{j} = \begin{pmatrix}
        \expval*{\hat{\Psi}(t, x, a)\hat{\Psi}^\dag(t, x, a)}_\beta & \expval*{\hat{\Psi}(t, x, a)\hat{\Psi}^\dag(t', y, b)}_\beta \\
        \expval*{\hat{\Psi}(t, x, a)\hat{\Psi}^\dag(t', y, b)}_\beta^\dag & \expval*{\hat{\Psi}(t', y, b)\hat{\Psi}^\dag(t', y, b)}_\beta
    \end{pmatrix},\quad
    C_\text{p} = \begin{pmatrix}
        \expval*{\hat{\Psi}(t, x, a)\hat{\Psi}^\dag(t, x, a)}_\beta & 0 \\
        0 & \expval*{\hat{\Psi}(t', y, b)\hat{\Psi}^\dag(t', y, b)}_\beta
    \end{pmatrix}.
\end{equation}
Substituting in the wavelet-transformed correlation functions $\expval*{\hat{\Psi}(t,x,a) \hat{\Psi}^\dagger(t',y,b)}_\beta$ derived in \cref{app:wavelet_correlators_fermionic}, the PRMI is expanded to second order since we are interested in the regime $\abs{x-y} \ll 1$.
To simplify the expressions we introduce the following wavelet coefficients
\begin{equation}
    M_{k,l}^{(i,j)}(a,b;\beta;m) =  
    \frac{1}{\pi a^{i}b^{j}}\int_0^{+\infty}\dd{p}p^{k}(p^2+m^2)^{l/2}(\,(\ii\hat{D})^{\underline{i}}\,\Tilde{w})^*_a(p)(\,(\ii\hat{D})^{\underline{j}}\,\Tilde{w})_b(p)
    \times\left\{\begin{array}{cc}
        1 & l\text{ even} \\
        f_\beta\big(\sqrt{p^2+m^2}\big) & l\text{ odd}
    \end{array}\right.,
    \label{eq:WaveletMCoeff}
\end{equation}
where $\hat{D}=\frac12(\hat{p}\cdot\hat{x}+\hat{x}\cdot\hat{p})$ is the dilation operator, $\hat{x}^{\underline{i}}=\hat{x}(\hat{x}-I)\cdots(\hat{x}-(i+1)I),\hat{x}^{\underline{0}}\equiv I$ is the falling factorial, and $f_\beta(\omega)=\tanh\tfrac12 \beta\omega$ comes from the fermionic statistics. 
Assuming a fixed scale $a=b=A$ and that the wavelet is (anti-)symmetric, this can be reduced using the properties of the $M$ symbols (see \cref{app:wavelet_transform}) to 
\begin{equation}
    I_2 =
    \log(\frac{16}{(1-A^2m^2{M_{0,-1}^{(0,0)}}^2)^2})-\frac{f_{tt}^\text{f}(\beta/A,Am)}{A^2}(t-t')^2-\frac{f_{xx}^\text{f}(\beta/A,Am)}{A^2}(x-y)^2\ ,
\end{equation}
where
\begin{align}
    f_{tt}^\text{f}(\beta/A,Am) &= 2\left(M_{0,2}^{(0,0)}-\frac{{M_{0,1}^{(0,0)}}^2-2A^2m^2M_{0,1}^{(0,0)}M_{0,-1}^{(0,0)}+A^2m^2}{1-A^2m^2{M_{0,-1}^{(0,0)}}^2}\right)\ ,
    \\
    f_{xx}^\text{f}(\beta/A,Am) &= 2\left(M_{2,0}^{(0,0)}-\frac{{M_{2,-1}^{(0,0)}}^2}{1-A^2m^2{M_{0,-1}^{(0,0)}}^2}\right)\ ,
\end{align}
and $M_{k,l}^{(i,j)}=M_{k,l}^{(i,j)}(1,1,\beta/A,Am)$.
In the massless case the coefficients in front of $(t-t')^2$ and $(x-y)^2$ are identical, i.e. $f_{tt}^\text{f}(\beta/A,0)=f_{xx}^\text{f}(\beta/A,0)$. And the mutual information around the coincidence follows the Euclidean Synge world function which can be obtained as
\begin{align}
    \frac12 \big((t - t')^2 + (x - y)^2\big)
    &= \frac{A^2}{2f_{xx}^\text{f}(\beta/A,0)} (\log16 - I_2)\ ,
\end{align}
where we arbitrarily decided to divide by $f_{xx}^{\text{f}}$ instead of $f_{tt}^{\text{f}}$. When $m\neq0$ the only difference is that $(t-t')^2$ is multiplied by $\frac{f_{tt}^\text{f}(\beta/A,Am)}{f_{xx}^\text{f}(\beta/A,Am)}$ which does not change the geometry as it can be removed by a co\"ordinate transform. Now the flat Euclidean metric can be obtained by
\begin{equation}
    g_{\mu\nu}^\text{f}
    = \frac{A^2}{2f_{xx}^\text{f}(\beta/A,Am)} \lim_{y^\rho\to x^\rho} \pdv{x^\mu} \pdv{y^\nu} I_2 ,
    \label{eq:metric_from_fermionic_prmi}
\end{equation}
with $x^0 = t, x^1 = x$ and $y^0 = t', y^1 = y$.
We now introduce the scale co\"ordinate $x^2 = a$ (resp. $y^2 = b$) by substituting the PRMI into \cref{eq:metric_from_fermionic_prmi} in order to derive the components of the metric tensor on this 3-dimensional manifold.
\begin{equation}
\dd{s}^2_\text{f} = \frac{A^2}{f_{xx}^\text{f}(\beta/A,Am)a^2}\Big(f_{tt}^\text{f}(\beta/a,am)\dd{t}^2+f_{xx}^\text{f}(\beta/a,am)\dd{x}^2+f_{aa}^\text{f}(\beta/a,am)\dd{a}^2\Big)\ ,
\end{equation}
with
\begin{equation}
    f_{aa}^\text{f}(\beta/A,Am) = 2M_{0,0}^{(1,1)}-\frac{A^2m^2}{2}\left(M_{0,-1}^{(0,1)}M_{0,-1}^{(1,0)}+2\frac{1+A^2m^2{M_{0,-1}^{(0,0)}}^2}{1-A^2m^2{M_{0,-1}^{(0,0)}}^2}({M_{0,-1}^{(0,1)}}^2+{M_{0,-1}^{(1,0)}}^2)\right)\ .
\end{equation}
In the massless, zero-temperature case the metric becomes
\begin{equation}
    \dd{s}^2_\text{f}
    = \frac{A^2}{a^2}( \dd{t}^2 + \dd{x}^2) + \frac{L_\text{f}^2}{a^2} \dd{a}^2\ ,
\end{equation}
which corresponds to the metric for the Poincaré half-space (the Riemannian counterpart of the anti-de Sitter spacetime). 
This is the first departure from the usual AdS$_3$/CFT$_2$ correspondence we observe: we obtain a Riemannian rather than Lorentzian geometry, a signature change also observed in the lattice wavelet approaches of \cite{qi2013,LeeQi:2016,SinghBrennen:2016,lee2017}.
Here $L_\text{f}$ is a wavelet dependent length scale given by
\begin{equation}
    L_\text{f}^2 
    = A^2 \frac{f_{aa}^\text{f}(\infty,0)}{f_{xx}^\text{f}(\infty,0)}
    = A^2\frac{-2W_0^{(2)}}{W_2^{(0)} - {W_1^{(0)}}^2}\ ,
\end{equation}
where $W_k^{(i)}$ are wavelet momentum and dilation moments defined as
\begin{equation}
\label{eq:wavelet_onesided_momentum_moment}
    W_k^{(i)}(a,b) 
    = \matrixel{w_a}
    {\abs{\hat{\bm{p}}}^k(\ii\hat{D})^i}{w_b}
    = \frac{1}
   {\pi}\int_0^{+\infty}p^{k}\Tilde{w}^*_a(p)
   ((\ii\hat{D})^i\tilde{w})_b(p)\dd{p}.
\end{equation}
To simplify certain comparisons and discussions we introduce a conformally equivalent metric $\Tilde{g}^\text{f}$ defined as
\begin{align}
    \Tilde{g}_{\mu\nu}^\text{f} &= 
    \frac12\lim_{y^\rho\to x^\rho} \pdv{x^\mu} \pdv{y^\nu}I_2((t, x, a);(t', y, b))\ ,
    \\
    \dd{\Tilde{s}}^2_\text{f} &=
    \frac{1}{a^2}\left(f_{tt}^\text{f}(\beta/a,am)\dd{t}^2+f_{xx}^\text{f}(\beta/a,am)\dd{x}^2+f_{aa}^\text{f}(\beta/a,am)\dd{a}^2\right)\ ,
    \label{eq:metricFerm}
\end{align}
and denote the corresponding geometric quantities with a tilde (e.g. $\Tilde{R}_\text{f}$ denotes the Ricci scalar of $\Tilde{g}^\text{f}$).
Although the coefficients $M_{k,l}^{(i,j)}$ in general do not have an analytic expression, the asymptotic behaviour of the geometry near the boundary ($a\to 0$) and deep in the bulk ($a\to\infty$) can still be investigated. Near the boundary the Ricci scalar $\Tilde{R}_\text{f}$ and metric components $\Tilde{g}_{\mu\mu}^\text{f}$ approach $3/W_0^{(2)}$ and $f_{\mu\mu}^\text{f}(\infty,0)/a^2$ respectively. Therefore, the resulting manifold is asymptotically AdS.

\begin{figure}[h]
    \centering
    \includegraphics[width=\columnwidth]{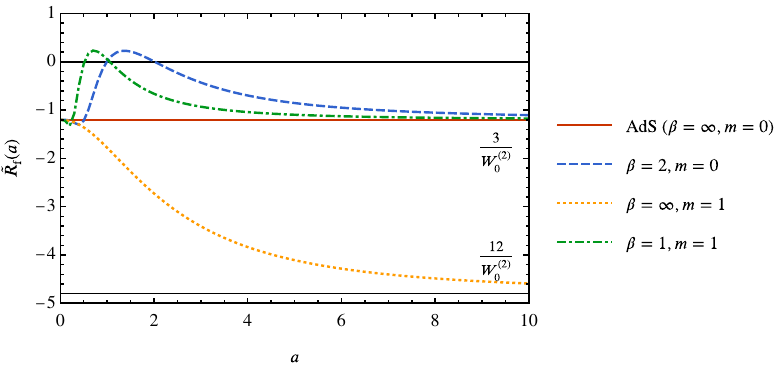}
    \caption{Plot of the Ricci scalar $\Tilde{R}_\text{f}$ as a function of the scale coordinate $a$ for the metric emergent from the $1+1$ dimensional Dirac fermion QFT, for different choices of mass and temperature. The wavelet used is the second Hermite wavelet (the second derivative of a Gaussian, i.e. the Mexican hat).}
    \label{fig:RicciScalarFermi}
\end{figure}

Deep in the bulk ($a\to\infty$) the metric components approach zero and the Ricci scalar approaches
\begin{equation}
    \lim_{a\to\infty}\Tilde{R}_\text{f}(a) = \left\{\begin{array}{cl}
        \frac{12}{W_0^{(2)}}, & \beta=\infty,\;m\neq0 \\
        \frac{3}{W_0^{(2)}}, & \beta\neq\infty,\;m=0,\text{ or }\beta=\infty,\;m=0
    \end{array}\right.
\end{equation}
indicating that near the center the manifold again resembles AdS with the either same or a different length scale than that of the asymptotic region (see \cref{fig:RicciScalarFermi}).

Because the nature of the emergent geometry depends on moments of the dilation operator and the momentum operator, it is appealing to seek a family of wavelets where these quantities can be tuned and/or extremized. One such family is affine group coherent state wavelets (see \cref{app:affine_group_wavelets}) defined in the momentum representation as 
\begin{equation}
    \tilde{w}_a(p)
    = \frac{\sgn{p}}{\sqrt{2\pi^{1/2}\sigma}}\frac{1}{\sqrt{\abs{p}}}
    \ee^{{-\frac{1}{2\sigma^2}(\log{\abs{p}}+\log{a})^2}}\ .
\end{equation}
These are coherent states with respect to the canonical pair $\hat{D}$ and $u=\log(|\hat{p}|)$, where $\sigma\in(0,+\infty)$ is the standard deviation of $u$.
The moment $W_0^{(2)}$, which is also minus the variance of the dilation operator for these wavelets, then reads
$W_0^{(2)} = -2/\sigma^2$.
By contrast, for Hermite wavelets indexed by $n\in\mathbb{Z}^+$
\begin{equation}
    w_n(x)=\frac{1}{\sqrt{\Gamma(n+\tfrac12)}}\dv[n]{x}\ee^{-\frac{x^2}{2}}\ ,
    \quad
    \Tilde{w}_n(p) = \sqrt{\frac{2\pi}{\Gamma(n+\tfrac12)}}(-\ii p)^n\ee^{-\frac{p^2}{2}}\ ,
\end{equation}
the moment attains the value $W_0^{(2)}=-(n+1/2)$.

%

\section{PRMI-derived metric for free boson}
\label{sec:prmi_metric_bosonic}
We next consider a free scalar field in a thermal state (a Gaussian state) and the Petz-Rényi mutual information with $\alpha=2$ given by
\begin{equation}
    I_2((t,x,a);(t',y,b))
    = -\frac{1}{2} \log \left( \frac{1+8(X_{1,2}P_{1,2} - \frac12(X_{1,1} P_{1,1} + X_{2,2} P_{2,2})) + 16 ((X_{1,2})^{2} - X_{1,1} X_{2,2})((P_{1,2})^{2} - P_{1,1} P_{2,2})}{(1 - 4 X_{1,1} P_{1,1})^2 (1 - 4 X_{2,2} P_{2,2})^2} \right)\ ,
    \label{eq:prmi_bosonic}
\end{equation}
where
\begin{align}
    X
    &= \frac12 \begin{pmatrix}
        \expval*{\{\hat{\phi}(t,x,a),\hat{\phi}(t,x,a)\}}_{\beta} & \expval*{\{\hat{\phi}(t,x,a),\hat{\phi}(t',y,b)\}}_{\beta}
        \\ 
        \expval*{\{\hat{\phi}(t,x,a),\hat{\phi}(t',y,b)\}}_{\beta} & \expval*{\{\hat{\phi}(t',y,b),\hat{\phi}(t',y,b)\}}_{\beta}
    \end{pmatrix}\ ,
    \\
    P
    &= \frac12 \begin{pmatrix}
        \expval*{\{\hat{\pi}(t,x,a),\hat{\pi}(t,x,a)\}}_{\beta} & \expval*{\{\hat{\pi}(t,x,a),\hat{\pi}(t',y,b)\}}_{\beta}
        \\ 
        \expval*{\{\hat{\pi}(t,x,a),\hat{\pi}(t',y,b)\}}_{\beta} & \expval*{\{\hat{\pi}(t',y,b),\hat{\pi}(t',y,b)\}}_{\beta}
    \end{pmatrix}\ .
\end{align}
In $(1+1)$ dimensions the expansions of wavelet-transformed correlation functions to second order around the $t=t', x=y$ coincidence reads
\begin{align}
    \nonumber X_{1,2}
    &= \frac12 \expval*{\{\hat{\phi}(t,x,a), \hat{\phi}(t',y,b)\}}_{\beta}
    = \frac12\Big(M_{0,-1}^{(0,0)}(a,b;\beta;m)-\frac12(M_{0,1}^{(0,0)}(a,b;\beta;m)(t-t')^2+M_{2,-1}^{(0,0)}(a,b;\beta;m)(x-y)^2)\Big)\ ,
    \\
    P_{1,2}
    &= \frac12 \expval*{\{\hat{\pi}(t,x,a), \hat{\pi}(t',y,b)\}}_{\beta}
    = \frac12\Big(M_{0,1}^{(0,0)}(a,b;\beta;m)-\frac12(M_{0,3}^{(0,0)}(a,b;\beta;m)(t-t')^2+M_{2,1}^{(0,0)}(a,b;\beta;m)(x-y)^2)\Big)\ ,
\end{align}
where $M_{k,l}^{(i,j)}(a,b;\beta;m)$ are defined in \cref{eq:WaveletMCoeff} with $f_\beta(\omega) = \coth\tfrac12\beta\omega$ coming from the bosonic statistics.
Fixing the scale $a=b=A$, the PRMI becomes
\begin{equation}
    I_2 = \log[(M_{0,-1}^{(0,0)}M_{0,1}^{(0,0)}-1)^2]+\frac{1}{A^2}\big(f_{tt}^\text{b}(\beta/A,Am)(t-t')^2+f_{xx}^\text{b}(\beta/A,Am)(x-y)^2\big)\ ,
\end{equation}
with
\begin{align}
    f_{tt}^\text{b}(\beta/A,Am) & = {M_{0,1}^{(0,0)}}^2+M_{0,-1}^{(0,0)}M_{0,3}^{(0,0)}\ ,
    \\
    f_{xx}^\text{b}(\beta/A,Am) & = M_{0,1}^{(0,0)}M_{2,-1}^{(0,0)}+M_{0,-1}^{(0,0)}M_{2,1}^{(0,0)}\ ,
\end{align}
where $M_{k,l}^{(i,j)}=M_{k,l}^{(i,j)}(1,1,\beta/A,Am)$. 
Again, in the massless case we have that $f_{tt}^\text{b}(\beta/A,0)=f_{xx}^\text{b}(\beta/A,0)$, and similarly the mutual information follows the Euclidean Synge world function around the coincidence. We again choose to explicitly divide by $f_{xx}^{\text{b}}$
\begin{equation}
    \frac12\big((t-t')^2 + (x-y)^2)\big)
    = \frac{A^2}{2f_{xx}^\text{b}(\beta/A,0)} \Big( I_2 - \frac12 \log \big( (1 - M_{0,-1}^{(0,0)} M_{0,1}^{(0,0)})^4 \big) \Big)\ ,
\end{equation}
which again leads to a constant rescaling of the time component for $m\neq0$. The flat Euclidean metric is obtained by
\begin{equation}
   g_{\mu\nu}^\text{b}
    = -\frac{A^2}{2f_{xx}^\text{b}(\beta/A,Am)} \lim_{y^\rho\to x^\rho} \pdv{x^\mu} \pdv{y^\nu} I_2((t, x, a);(t', y, b))\ ,
    \label{eq:metric_from_bosonic_prmi}
\end{equation}
with $x^0 = t, x^1 = x$ and $y^0 = t', y^1 = y$ as before. 

As before we introduce the scale co\"ordinate $x^2 = a$ (resp. $y^2 = b$) and use \cref{eq:metric_from_bosonic_prmi} to derive the metric. The result is
\begin{equation}
    \dd{s}^2_\text{b} = \frac{A^2}{f_{xx}^\text{b}(\beta/A,Am)a^2}\Big(f_{tt}^\text{b}(\beta/a,am)\dd{t}^2+f_{xx}^\text{b}(\beta/a,am)\dd{x}^2+f_{aa}^\text{b}(\beta/a,am)\dd{a}^2\Big)\ ,
    \label{metricbos}
\end{equation}
where
\begin{align}
    f_{aa}^\text{b}(\beta/A,Am) = \frac12 \big({M_{0,1}^{(0,0)}}^2 (M_{0,-1}^{(0,1)}-M_{0,-1}^{(1,0)})^2+&{M_{0,-1}^{(0,0)}}^2 (M_{0,1}^{(0,1)}-M_{0,1}^{(1,0)})^2+2 M_{0,1}^{(0,1)} M_{0,-1}^{(1,0)}\\
    &\qquad+2 M_{0,-1}^{(0,1)} M_{0,1}^{(1,0)}+2 M_{0,1}^{(0,0)} M_{0,-1}^{(1,1)}+2 M_{0,-1}^{(0,0)} M_{0,1}^{(1,1)}\big)\ .\nonumber
\end{align}
For $\beta = \infty$ and $m = 0$ the metric reduces to
\begin{equation}
    \dd{s}^2_\text{b} = \frac{A^2}{a^2}(\dd{t}^2+\dd{x}^2) + \frac{L^2_\text{b}}{a^2}\dd{a}^2\ ,
\end{equation}
where $L_\text{b}$ is the wavelet-dependent length scale.
The metric obtained is again that of a Poincaré half-space which is the Riemannian counterpart of the AdS spacetime.
This time however the expression for the length scale is more involved when written in terms of the wavelet moments $W_k^{(i)}$ (see \cref{eq:wavelet_onesided_momentum_moment}):
\begin{equation}
    L_\text{b}^2 
    = \frac{ A^2 }{ {W_1^{(0)}}^2 + W_{-1}^{(0)} W_3^{(0)} } \big( 2 {W_{-1}^{(0)}}^2 ( W_1^{(1)} + \tfrac12 W_1^{(0)})^2 + 2 {W_1^{(0)}}^2 ( W_{-1}^{(1)} - \tfrac12 W_{-1}^{(0)} )^2 - 2 W_{-1}^{(1)} W_1^{(1)} - W_{-1}^{(0)} W_1^{(2)} - W_1^{(0)} W_{-1}^{(2)}\big)
    \label{eq:bosonic_radius_of_curvature}
\end{equation}
As before, we introduce a conformally equivalent metric $\Tilde{g}^\text{b}$ as
\begin{align}
    \Tilde{g}_{\mu\nu}^\text{b} &=
    -\frac12\lim_{y^\rho\to x^\rho} \pdv{x^\mu} \pdv{y^\nu} I_2((t, x, a);(t', y, b))\ ,
    \\
    \dd{\Tilde{s}}^2_\text{b} &= 
    \frac{1}{a^2}\Big(f_{tt}^\text{b}(\beta/a,am)\dd{t}^2+f_{xx}^\text{b}(\beta/a,am)\dd{x}^2+f_{aa}^\text{b}(\beta/a,am)\dd{a}^2\Big)\ ,
    \label{eq:metricBosonic}
\end{align}
and denote the corresponding geometric quantities with a tilde. Note the minus sign in the expression for the metric relative to \cref{eq:metricFerm} for the fermionic theory.
This is due to the Pauli exclusion principle: in the fermionic case, the two-point correlations near coincidence decrease with decreasing separation, while they increase for the bosonic case.
This results in opposite signs in the quadratic expansions for the respective PRMIs.

Here again the coefficients $M_{k,l}^{(i,j)}$ do not have an analytic expression but the asymptotic behaviour of the geometry can still be studied.
Near the boundary the Ricci scalar $\Tilde{R}_\text{b}$ and metric components $\Tilde{g}_{\mu\mu}^\text{b}$ limit to $-6 /f_{aa}^\text{b}(\infty,0) $ and $f_{\mu\mu}^\text{b}(\infty,0)/a^2$ respectively.
The manifold is again asymptotically AdS.
In the massive case we find that deep in the bulk the metric components vanish except for
\begin{equation}
    \lim_{a\to\infty}\Tilde{g}_{tt}^\text{b}(a)= 2m^2\coth^2\frac{m\beta}{2}\ ,
\end{equation}
and the Ricci scalar limits to
\begin{equation}
    \lim_{a\to\infty}\Tilde{R}_\text{b}(a) = \frac{\tanh^2\frac{m\beta}{2}}{W_0^{(2)}}\ .
\end{equation}
The massless case is qualitatively different. 
The Ricci scalar vanishes deep in the bulk and the metric components approach constant non-zero values
\begin{align}
    \lim_{a\to\infty}\Tilde{g}_{tt}^\text{b}(a)=\lim_{a\to\infty}\Tilde{g}_{xx}^\text{b}(a) &= \frac{4}{\beta^2}(1+W^{(0)}_{-2}W^{(0)}_2)\ , \\
    \lim_{a\to\infty}\Tilde{g}_{aa}^\text{b}(a) &= \frac{4}{\beta^2}(2W_{-2}^{(0)}-W_{-2}^{(2)}-W_{-2}^{(0)}W_0^{(2)})\ .
\end{align}
Therefore, in the massless thermal case the deep bulk geometry appears to be flat. This emergent IR Minkowski region represents a departure from standard holographic RG flow scenarios. Geometries such as this arise when there is a deformation of the 2d CFT (a breaking of the conformal symmetry) driven by vacuum expectation values of irrelevant operators. Such metrics have been studied in the context of exotic irrelevant deformations in gauged supergravity \cite{astrakhantsev2025}.

\begin{figure}[h]
    \centering
    \includegraphics[width=\columnwidth]{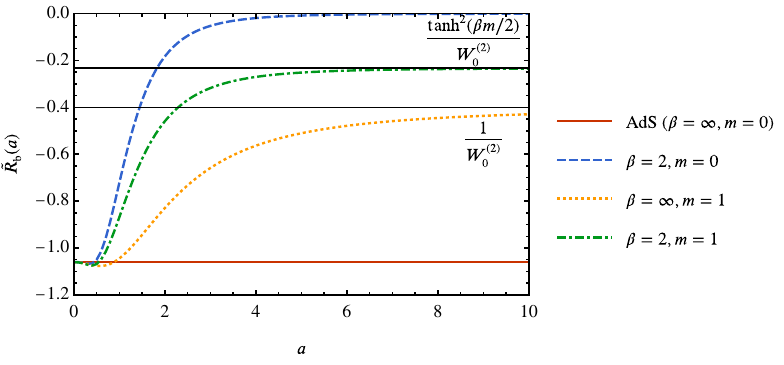}
    \caption{Plot of Ricci scalar $\Tilde{R}_\text{b}$ as a function of the scale co\"ordinate for the metric emergent from the free $1+1$ dimensional scalar bosonic QFT for different choices of mass and temperature. The wavelet used is the second Hermite wavelet (second derivative of a Gaussian or Mexican hat)}
    \label{fig:RicciScalarBosonic}
\end{figure}

\subsection{The bulk matter content}
The behaviour of the Ricci scalar in \cref{fig:RicciScalarBosonic} is reminiscent of domain wall solutions of scalar-tensor Einstein theory \cite{Charmousis2017}
\begin{equation}
    S[g_{\mu\nu},\phi] = \int \dd[3]{x}\sqrt{|g|}\left[R-\frac12 g^{\mu\nu}\partial_\mu\phi\;\partial_\nu\phi-V(\phi)\right],
\end{equation}
where the scalar field assumes a constant value at the boundary and a different constant value deep in the bulk. As a result the geometry interpolates between two AdS spaces with distinct length scales derived from the value of the scalar field. The metric in \cref{eq:metricBosonic} can be cast into the form of (asymmetrically) warped (Euclidean) AdS space
\begin{equation}
    \dd{\Tilde{s}_{\text{b}}}^2 = \ee^{2\mathcal{A}(r)}\dd{t}^2+\ee^{2\mathcal{B}(r)}\dd{x}^2+\dd{r}^2,
\end{equation}
where the new coordinate $r$ is defined as $r(a) = \int^a\sqrt{\tilde{g}_{aa}(a')}\dd{a'}$. Assuming, as in \cite{Charmousis2017}, that $\phi=\phi(r)$ one can use the Einstein field equations and the Klein-Gordon equation to derive constraints for the two warp factors $\mathcal{A}(r)$ and $\mathcal{B}(r)$. 
We tested one of these constraints
\begin{equation}
-(\mathcal{A}''(r)+\mathcal{B}''(r))-(\mathcal{B}'(r)-\mathcal{A}'(r))^2=2\phi'(r)^2\geq 0, 
\end{equation}
where $'\equiv \partial_r$,
numerically for several choices of mass, temperature and wavelet function and observed that our metric \cref{eq:metricBosonic} did not satisfy it for any choice of potential $V$. This leads us to conclude that our bulk metric cannot be interpreted as a solution of a simple gravity-matter model, and further analysis beyond the scope of this work is necessary. This is perhaps unsurprising, since domain wall-like solutions in holographic theories typically arise from consistent truncations of supergravity that retain substantial field content, including multiple scalars with self-interaction potentials, gauge fields, and/or higher-spin fields, going well beyond the minimal Einstein-scalar system we have tested against.

\subsection{Free scalar field in 1+d}
\label{sec:prmi_metric_bosonic_1plusd}

These results can be readily generalised to the case of a free scalar field in $1 + d$ dimensions, as
\Cref{eq:prmi_bosonic} remains valid for $d>1$.
By using wavelets that are isotropic and therefore depend only on the spatial distance, the Fourier transform can be reduced to a Hankel transform and the correlation functions can be expanded around coincidence as (see \cref{app:wavelet_correlators_bosonic})
\begin{align}
    \nonumber X_{1,2}
    &= \frac12 \expval*{\{\hat{\phi}(t,\bm{x},a), \hat{\phi}(t',\bm{y},b)\}}_{\beta}
    = \frac12\Big(M_{0,-1}^{(0,0)}(a,b;\beta;m)-\frac12(M_{0,1}^{(0,0)}(a,b;\beta;m)(t-t')^2+\tfrac{1}{d}M_{2,-1}^{(0,0)}(a,b;\beta;m)|\bm{x}-\bm{y}|^2)\Big)
    \\
    P_{1,2}
    &= \frac12 \expval*{\{\hat{\pi}(t,\bm{x},a), \hat{\pi}(t',\bm{y},b)\}}_{\beta}
    = \frac12\Big(M_{0,1}^{(0,0)}(a,b;\beta;m)-\frac12(M_{0,3}^{(0,0)}(a,b;\beta;m)(t-t')^2+\tfrac{1}{d}M_{2,1}^{(0,0)}(a,b;\beta;m)|\bm{x}-\bm{y}|^2)\Big),
    \label{eq:bosonic_wavelet_correlator_manyd}
\end{align}
Substituting these expansions into \cref{eq:prmi_bosonic} will result in a result similar to \cref{eq:metric_from_bosonic_prmi}, except that the spatial components of the metric tensor are rescaled by the dimension
\begin{equation}
    g_{ij}^\text{b} = \frac{1}{d}\frac{A^2}{\big({M_{0,1}^{(0,0)}}^2 + M_{0,3}^{(0,0)} M_{0,-1}^{(0,0)}\big)a^2}f_{xx}^\text{b}(\beta/a,am)\delta_{ij},
\end{equation}
and the $M$ coefficients now follow the higher dimensional definition \cref{eq:WaveletMDefinition_Appendix}. This, however, does not meaningfully impact the computations of the previous section, and the Ricci scalar will still have the same symbolic form when written in terms of the $M$, which share similar limiting behaviour as seen in \cref{fig:RicciScalarBosonic}.


\section{Discussion and Conclusion}
\label{sec:discussion_conclusion}

We have developed a wavelet-based approach to reverse holography that constructs higher dimensional bulk geometries from lower dimensional field theory data. We began by deriving an analogue to the main result of \textcite{saravani2016} for a Dirac field, demonstrating several inconsistencies when applying the method directly to wavelet-transformed correlation functions, including complex metric tensors and a disagreement between particle statistics based on the type of geometry. Our central technical advance is replacing two-point correlation functions with a more robust entropic measure, the Petz-Rényi mutual information, and treating it as the fundamental field theoretic quantity. This basis-independent measure resolves inconsistencies between bosonic and fermionic theories and produces geometries that are asymptotically AdS in the UV, with IR behavior controlled by mass and temperature scales of the `boundary' theory.

Specifically, we considered the Petz-Rényi mutual information with $\alpha=2$ and derived metric tensors for the wavelet-transformed bosonic and fermionic theories in flat space.
We find that with our approach, the vacuum and thermal states of the scalar and Dirac fields produce a similar type of geometry; asymmetrically warped Euclidean AdS spacetime, which becomes Euclidean AdS for the vacuum state when the two theories are massless (and thus conformal).
However, many features of this wavelet-based approach to holography differ from that of traditional AdS-CFT holography, since in the CFT regime the geometry obtained here is Riemannian and not Lorentzian. This change in signature has been observed before in the context of the exact holographic mapping using discrete wavelets \cite{qi2013,LeeQi:2016,SinghBrennen:2016,lee2017}. 
In those examples, like here, the wavelet transform was performed over spatial degrees of freedom only. Unlike our method however, the metric there was inferred from the distance between bulk points, which was calculated by assuming an exponential decay of mutual information between points on a Euclidean bulk space and matching that decay to the computed values. This is fundamentally different than our approach, since there a bulk metric is assumed to exist a priori, whereas we compute PRMI between boundary field points without presupposing a bulk spacetime structure.

Euclidean metrics have also been derived for QFTs using the tools of information geometry \cite{junior2017, erdmenger2020}.
In this approach, the co\"ordinates are input parameters to the theory (like mass or temperature) and the Fisher metric is computed from the path integral expectation value of derivatives of the action with respect to those co\"ordinates. Such a calculation depends on the global equilibrium states. In contrast, our metric is derived from the mutual information between two local reduced states of the field as resolved by a pair of detectors.
As such it is more directly related to spacetime as it is derived from varying the position of one of the detectors rather than parameters of the theory \cite{erdmenger2020}.

There are several advantages to the wavelet approach to holography.
As we have seen, it can be straightforwardly extended to non-CFTs such as massive QFTs and the formalism allows for extensions to QFTs with broken translational symmetry, e.g. the presence of mass defects.
Secondly, the presented method relies only on local quantities, in contrast to e.g. cMERA \cite{Nozaki2012} where the scale component of the metric is obtained from the Bures distance which considers the total state.
Lastly, the wavelet picture offers an interesting interpretation for the bulk manifold.
One can interpret the points in the manifold as corresponding to detector configurations where each configuration is labelled by time, position, and resolution (scale).
The cutoff scale then corresponds to the finest resolution of the detector.

As shown in \cref{fig:RicciScalarBosonic}, the Ricci scalar asymptotes to two different constant values in the centre and near the boundary of the bulk spacetime.
We have examined the field content of the bulk for an effective matter-gravity model and observed that the resulting metric is not sourced by a scalar field. This suggests that one might need to employ higher curvature theories or more exotic field content to interpret the wavelet bulk as geometry sourced by matter.

This study opens several interesting paths for future investigations.
In our construction, the bulk metric has the AdS length scale controlled by the choice of wavelet and reference scale, so it will be of further interest to examine how properties of the wavelet are mapped to properties of the RG flow.
On a fundamental level, this work makes a step towards understanding the relationship between QFT, information theory, and geometry.
Specifically, how one could infer distance from an information-theoretic quantity that is independent of the choice of basis for observables.
The method and entropic quantity (PRMI with $\alpha = 2$) we present in this paper is just one of many possibilities which suggests the question: is there a criterion that might fix a method for extracting the metric from mutual information?
Or, which is the most natural method and entropic quantity for this? 
To resolve these questions it might be useful to consider the Hadamard condition \cite{fulling1978} which states that two-point functions of Hadamard states on curved background follow a certain expansion in geodesic distance around the coincidence. 
It is unclear how to apply the Hadamard condition in the context of wavelets which opens an interesting avenue of research.

\begin{acknowledgments}
This work was supported by the Australian Research Council (ARC) Discovery Project scheme (ARC project no. DP200102152).
\v{S}V, DJG, and GKB acknowledge support from the ARC Centre of Excellence for Engineered Quantum Systems (ARC project no. CE170100009). FS acknowledges support from the ARC Discovery Project scheme (ARC project no. DP200102152).
DGL, NF, and NCM acknowledge support from the ARC Centre of Excellence for Quantum Computation and Communication Technology (ARC project no. CE170100012).
NCM acknowledges support from the ARC Future Fellowship scheme (ARC project no. FT230100571).
AK acknowledges support from the National Science and Engineering Research Council of Canada (NSERC) and the National Research Council of Canada (NRC).
\v{S}V and DJG acknowledge support from the Sydney Quantum Academy, Sydney, Australia.
\end{acknowledgments}

\bibliography{main}

@misc{qi2013,
      title={Exact holographic mapping and emergent space-time geometry}, 
      author={Xiao-Liang Qi},
      year={2013},
      eprint={1309.6282},
      archivePrefix={arXiv},
      primaryClass={hep-th},
      url={https://arxiv.org/abs/1309.6282}, 
}

@article{susskind1995,
  title = {The {{World}} as a {{Hologram}}},
  author = {Susskind, L.},
  year = 1995,
  month = nov,
  journal = {Journal of Mathematical Physics},
  volume = {36},
  number = {11},
  eprint = {hep-th/9409089},
  pages = {6377--6396},
  issn = {0022-2488, 1089-7658},
  doi = {10.1063/1.531249},
  urldate = {2020-04-21},
  archiveprefix = {arXiv},
}

@article{hooft1993,
  title = {Dimensional {{Reduction}} in {{Quantum Gravity}}},
  author = {'t Hooft, G.},
  year = 1993,
  month = mar,
  journal = {arXiv:gr-qc/9310026},
  eprint = {gr-qc/9310026},
  urldate = {2020-04-21},
  archiveprefix = {arXiv},
}

@article{Lee2017,
  title = {Generalized exact holographic mapping with wavelets},
  author = {Lee, Ching Hua},
  journal = {Phys. Rev. B},
  volume = {96},
  issue = {24},
  pages = {245103},
  numpages = {13},
  year = {2017},
  month = {Dec},
  publisher = {American Physical Society},
  doi = {10.1103/PhysRevB.96.245103},
  url = {https://link.aps.org/doi/10.1103/PhysRevB.96.245103}
}

@misc{astrakhantsev2025,
      title={Zoo of flows in a 3d gauged supergravity with periodic potential}, 
      author={Astrakhantsev, Lev and Golubtsova, Anastasia A. and Podoinitsyn, Mikhail A.},
      year={2025},
      eprint={2511.21558},
      archivePrefix={arXiv},
      primaryClass={hep-th},
      url={https://arxiv.org/abs/2511.21558}, 
}

@inproceedings{altaisky2012,
  title={On wavelet transform in {Minkowski} space},
  author={Mikhail V. Altaisky and Natalia E. Kaputkina},
  booktitle={2012 Proceedings of the International Conference Days on Diffraction},
  year={2012},
  pages={17-20},
  url={https://api.semanticscholar.org/CorpusID:32665195}
}

@article{ryu2006c,
  title = {Holographic Derivation of Entanglement Entropy from {{AdS}}/{{CFT}}},
  author = {Ryu, Shinsei and Takayanagi, Tadashi},
  year = 2006,
  journal = {Phys. Rev. Lett.},
  volume = {96},
  pages = {181602},
  doi = {10.1103/PhysRevLett.96.181602}
}

@article{buchel2005b,
  title = {Coupling Constant Dependence of the Shear Viscosity in {{N}} = 4 Supersymmetric {{Yang}}--{{Mills}} Theory},
  author = {Buchel, Alex and Liu, James T. and Starinets, Andrei O.},
  year = 2005,
  month = feb,
  journal = {Nuclear Physics B},
  volume = {707},
  number = {1-2},
  pages = {56--68},
  issn = {05503213},
  doi = {10.1016/j.nuclphysb.2004.11.055},
  urldate = {2025-11-16},
  langid = {english},
}

@article{kovtun2005b,
  title = {Viscosity in {{Strongly Interacting Quantum Field Theories}} from {{Black Hole Physics}}},
  author = {Kovtun, P. K. and Son, D. T. and Starinets, A. O.},
  year = 2005,
  month = mar,
  journal = {Phys. Rev. Lett.},
  volume = {94},
  number = {11},
  pages = {111601},
  issn = {0031-9007, 1079-7114},
  doi = {10.1103/PhysRevLett.94.111601},
  urldate = {2025-11-16},
  copyright = {http://link.aps.org/licenses/aps-default-license},
  langid = {english},
}

@article{altaisky2013,
  title = {Continuous Wavelet Transform in Quantum Field Theory},
  author = {Altaisky, M. V. and Kaputkina, N. E.},
  date = {2013-07-10},
  year = {2013},
  journal = {Physical Review D},
  shortjournal = {Phys. Rev. D},
  volume = {88},
  number = {2},
  pages = {025015},
  issn = {1550-7998, 1550-2368},
  doi = {10.1103/PhysRevD.88.025015},
  url = {https://link.aps.org/doi/10.1103/PhysRevD.88.025015},
  urldate = {2023-02-22},
  langid = {english}
}

@article{altaisky2022,
  title = {Can {{Our Spacetime Emerge}} from {{Anti}}--de {{Sitter Space}}?},
  author = {Altaisky, M. V. and Raj, R.},
  year = 2022,
  month = aug,
  journal = {Physics of Particles and Nuclei Letters},
  volume = {19},
  number = {4},
  pages = {313--316},
  issn = {1547-4771, 1531-8567},
  doi = {10.1134/S1547477122040033},
  urldate = {2025-10-16},
  langid = {english}
}

@Article{casini2018,
author={Casini, Horacio
and Medina, Raimel
and Landea, Ignacio Salazar
and Torroba, Gonzalo},
title={Renyi relative entropies and renormalization group flows},
journal={Journal of High Energy Physics},
year={2018},
month={Sep},
day={28},
volume={2018},
number={9},
pages={166},
issn={1029-8479},
doi={10.1007/JHEP09(2018)166},
url={https://doi.org/10.1007/JHEP09(2018)166}
}

@article{erdmenger2020,
  title = {Information Geometry in Quantum Field Theory: Lessons from Simple Examples},
  shorttitle = {Information Geometry in Quantum Field Theory},
  author = {Erdmenger, Johanna and Grosvenor, Kevin and Jefferson, Ro},
  year = 2020,
  month = may,
  journal = {SciPost Physics},
  volume = {8},
  number = {5},
  pages = {073},
  issn = {2542-4653},
  doi = {10.21468/SciPostPhys.8.5.073},
  urldate = {2025-10-16},
  langid = {english}
}

@article{fulling1978,
  title = {Singularity Structure of the Two-Point Function in Quantum Field Theory in Curved Spacetime},
  author = {Fulling, Stephen A. and Sweeny, Mark and Wald, Robert M.},
  date = {1978-10-01},
  year = {1978},
  month = {Oct},
  day = {1},
  journal = {Communications in Mathematical Physics},
  shortjournal = {Commun.Math. Phys.},
  volume = {63},
  number = {3},
  pages = {257--264},
  issn = {1432-0916},
  doi = {10.1007/BF01196934},
  url = {https://doi.org/10.1007/BF01196934},
  urldate = {2025-04-09},
  langid = {english}
}

@article{george2022,
  title = {Entanglement in quantum field theory via wavelet representations},
  author = {George, Daniel J. and Sanders, Yuval R. and Bagherimehrab, Mohsen and Sanders, Barry C. and Brennen, Gavin K.},
  journal = {Phys. Rev. D},
  volume = {106},
  issue = {3},
  pages = {036025},
  numpages = {19},
  year = {2022},
  month = {Aug},
  publisher = {American Physical Society},
  doi = {10.1103/PhysRevD.106.036025},
  url = {https://link.aps.org/doi/10.1103/PhysRevD.106.036025}
}

@INPROCEEDINGS{gorodnitskiy2011,
  author={Gorodnitskiy, Evgeny A. and Perel, Maria V.},
  booktitle={Proceedings of the International Conference Days on Diffraction 2011}, 
  title={The {Poincaré} wavelet transform: Implementation and interpretation}, 
  year={2011},
  volume={},
  number={},
  pages={72-77},
  doi={10.1109/DD.2011.6094368}}

@article{Haegeman:2013,
  title = {Entanglement Renormalization for Quantum Fields in Real Space},
  author = {Haegeman, Jutho and Osborne, Tobias J. and Verschelde, Henri and Verstraete, Frank},
  journal = {Phys. Rev. Lett.},
  volume = {110},
  issue = {10},
  pages = {100402},
  numpages = {5},
  year = {2013},
  month = {Mar},
  publisher = {American Physical Society},
  doi = {10.1103/PhysRevLett.110.100402},
  url = {https://link.aps.org/doi/10.1103/PhysRevLett.110.100402}
}

@article{junior2017,
  title = {Geometry of the {{Shannon}} Mutual Information in Continuum {{QFT}}},
  author = {Junior, David R. and Oxman, Luis E.},
  year = 2017,
  month = jun,
  journal = {Physical Review D},
  volume = {95},
  number = {12},
  eprint = {1704.02040},
  primaryclass = {hep-th},
  pages = {125005},
  issn = {2470-0010, 2470-0029},
  doi = {10.1103/PhysRevD.95.125005},
  urldate = {2025-10-16},
  archiveprefix = {arXiv}
}

@article{kempf2021,
  author  = {Kempf, A},
  title   = {Replacing the Notion of Spacetime Distance by the Notion of Correlation},
  journal = {Frontiers in Physics},
  year    = {2021},
  volume  = {9},
  pages   = {655857},
  doi     = {10.3389/fphy.2021.655857}
}

@article{kudler-flam2023,
  title = {R\'enyi Mutual Information in Quantum Field Theory},
  author = {Kudler-Flam, Jonah},
  journal = {Phys. Rev. Lett.},
  volume = {130},
  issue = {2},
  pages = {021603},
  numpages = {6},
  year = {2023},
  month = {Jan},
  publisher = {American Physical Society},
  doi = {10.1103/PhysRevLett.130.021603},
  url = {https://link.aps.org/doi/10.1103/PhysRevLett.130.021603}
}

@article{LeeQi:2016,
  title = {Exact holographic mapping in free fermion systems},
  author = {Lee, Ching Hua and Qi, Xiao-Liang},
  journal = {Phys. Rev. B},
  volume = {93},
  issue = {3},
  pages = {035112},
  numpages = {27},
  year = {2016},
  month = {Jan},
  publisher = {American Physical Society},
  doi = {10.1103/PhysRevB.93.035112},
  url = {https://link.aps.org/doi/10.1103/PhysRevB.93.035112}
}

@Article{Maldacena1999,
author={Maldacena, Juan},
title={The Large-N Limit of Superconformal Field Theories and Supergravity},
journal={International Journal of Theoretical Physics},
year={1999},
month={Apr},
day={01},
volume={38},
number={4},
pages={1113-1133},
issn={1572-9575},
doi={10.1023/A:1026654312961},
url={https://doi.org/10.1023/A:1026654312961}
}

@Article{McMahon2020,
author={McMahon, Nathan A.
and Singh, Sukhbinder
and Brennen, Gavin K.},
title={A holographic duality from lifted tensor networks},
journal={npj Quantum Information},
year={2020},
month={Apr},
day={24},
volume={6},
number={1},
pages={36},
issn={2056-6387},
doi={10.1038/s41534-020-0255-7},
url={https://doi.org/10.1038/s41534-020-0255-7}
}

@Article{Nozaki2012,
author={Nozaki, Masahiro
and Ryu, Shinsei
and Takayanagi, Tadashi},
title={Holographic geometry of entanglement renormalization in quantum field theories},
journal={Journal of High Energy Physics},
year={2012},
month={Oct},
day={30},
volume={2012},
number={10},
pages={193},
issn={1029-8479},
doi={10.1007/JHEP10(2012)193},
url={https://doi.org/10.1007/JHEP10(2012)193}
}

@article{pastawski2015,
  title = {Holographic Quantum Error-Correcting Codes: {{Toy}} Models for the Bulk/Boundary Correspondence},
  shorttitle = {Holographic Quantum Error-Correcting Codes},
  author = {Pastawski, Fernando and Yoshida, Beni and Harlow, Daniel and Preskill, John},
  year = {2015},
  month = {Jun},
  journal = {Journal of High Energy Physics},
  volume = {2015},
  number = {6},
  eprint = {1503.06237},
  primaryclass = {hep-th},
  pages = {149},
  issn = {1029-8479},
  doi = {10.1007/JHEP06(2015)149},
  urldate = {2025-11-18},
  archiveprefix = {arXiv}
}

@article{perche2022,
  title = {Geometry of spacetime from quantum measurements},
  author = {Perche, T. Rick and Mart\'{\i}n-Mart\'{\i}nez, Eduardo},
  journal = {Phys. Rev. D},
  volume = {105},
  issue = {6},
  pages = {066011},
  numpages = {21},
  year = {2022},
  month = {Mar},
  publisher = {American Physical Society},
  doi = {10.1103/PhysRevD.105.066011},
  url = {https://link.aps.org/doi/10.1103/PhysRevD.105.066011}
}

@article{PETZ198657,
title = {Quasi-entropies for finite quantum systems},
journal = {Reports on Mathematical Physics},
volume = {23},
number = {1},
pages = {57-65},
year = {1986},
issn = {0034-4877},
doi = {https://doi.org/10.1016/0034-4877(86)90067-4},
url = {https://www.sciencedirect.com/science/article/pii/0034487786900674},
author = {D\'{e}nes Petz}
}

@article{salvio2025,
  title = {Introduction to {{Thermal Field Theory}}: {{From First Principles}} to {{Applications}}},
  shorttitle = {Introduction to {{Thermal Field Theory}}},
  author = {Salvio, Alberto},
  year = {2025},
  month = {Jan},
  journal = {Universe},
  volume = {11},
  number = {1},
  eprint = {2411.02498},
  primaryclass = {hep-ph},
  pages = {16},
  issn = {2218-1997},
  doi = {10.3390/universe11010016},
  urldate = {2025-09-04},
  archiveprefix = {arXiv}
}

@article{saravani2016,
  title = {Spacetime curvature in terms of scalar field propagators},
  author = {Saravani, Mehdi and Aslanbeigi, Siavash and Kempf, Achim},
  journal = {Phys. Rev. D},
  volume = {93},
  issue = {4},
  pages = {045026},
  numpages = {13},
  year = {2016},
  month = {Feb},
  publisher = {American Physical Society},
  doi = {10.1103/PhysRevD.93.045026},
  url = {https://link.aps.org/doi/10.1103/PhysRevD.93.045026}
}

@misc{SinghBrennen:2016,
      title={Holographic Construction of Quantum Field Theory using Wavelets}, 
      author={Sukhwinder Singh and Gavin K. Brennen},
      year={2016},
      eprint={1606.05068},
      archivePrefix={arXiv},
      primaryClass={quant-ph},
      url={https://arxiv.org/abs/1606.05068}, 
}

@article{Singh:2018,
  title = {Holographic spin networks from tensor network states},
  author = {Singh, Sukhwinder and McMahon, Nathan A. and Brennen, Gavin K.},
  journal = {Phys. Rev. D},
  volume = {97},
  issue = {2},
  pages = {026013},
  numpages = {22},
  year = {2018},
  month = {Jan},
  publisher = {American Physical Society},
  doi = {10.1103/PhysRevD.97.026013},
  url = {https://link.aps.org/doi/10.1103/PhysRevD.97.026013}
}

@article{Swingle:2012,
  title = {Entanglement renormalization and holography},
  author = {Swingle, Brian},
  journal = {Phys. Rev. D},
  volume = {86},
  issue = {6},
  pages = {065007},
  numpages = {8},
  year = {2012},
  month = {Sep},
  publisher = {American Physical Society},
  doi = {10.1103/PhysRevD.86.065007},
  url = {https://link.aps.org/doi/10.1103/PhysRevD.86.065007}
}

@article{Vidal:2007,
  title = {Entanglement Renormalization},
  author = {Vidal, G.},
  journal = {Phys. Rev. Lett.},
  volume = {99},
  issue = {22},
  pages = {220405},
  numpages = {4},
  year = {2007},
  month = {Nov},
  publisher = {American Physical Society},
  doi = {10.1103/PhysRevLett.99.220405},
  url = {https://link.aps.org/doi/10.1103/PhysRevLett.99.220405}
}

@article{Witten:1998qj,
    author = "Witten, Edward",
    title = "{Anti de Sitter space and holography}",
    eprint = "hep-th/9802150",
    archivePrefix = "arXiv",
    reportNumber = "IASSNS-HEP-98-15",
    doi = "10.4310/ATMP.1998.v2.n2.a2",
    journal = "Adv. Theor. Math. Phys.",
    volume = "2",
    pages = "253--291",
    year = "1998"
}

@Article{Charmousis2017,
author={Charmousis, Christos
and Kiritsis, Elias
and Nitti, Francesco},
title={Holographic self-tuning of the cosmological constant},
journal={Journal of High Energy Physics},
year={2017},
month={Sep},
day={08},
volume={2017},
number={9},
pages={31},
issn={1029-8479},
doi={10.1007/JHEP09(2017)031},
url={https://doi.org/10.1007/JHEP09(2017)031}
}


\appendix

%

\section{Continuous wavelet transform}
\label{app:wavelet_transform}

The continuous wavelet transform on $L^2(\mathbb{R})$ is defined as
\begin{equation}
    \phi(x,a) 
    = \braket{w_{a,x}}{\phi} 
    = \int_{-\infty}^{+\infty}w^*_{a,x}(x')\phi(x')\dd{x'},
    \label{eq:continuous_wavelet_transform_1d}
\end{equation}
where $w_{a,x}(x')=a^{-1/2}w(\frac{x'-x}{a})$ and $a>0$. 
We extend this definition to $d$ dimensions as
\begin{equation}
    \phi(\bm{x},a,\theta) 
    = \braket{w_{a,\theta,\bm{x}}}{\phi} 
    = \int_{\mathbb{R}^d}w^*_{a,\theta,\bm{x}}(\bm{x}')\phi(\bm{x}')\dd[d]{\bm{x}'} 
    = \int_{\mathbb{R}^d}\ee^{-\ii\bm{p}\cdot\bm{x}}\Tilde{w}_{a,\theta}(\bm{p})\Tilde{\phi}(\bm{p})\frac{\dd[d]{\bm{p}}}{(2\pi)^d},
    \label{eq:continuous_wavelet_transform_manyd}
\end{equation}
where $w$ is the wavelet function, $a>0$ is the wavelet scale, $\theta \in SO(d)$ is the orientation of the wavelet function, and
\begin{equation}
    w_{a,\theta,\bm{x}}(\bm{x}') 
    = \frac{1}{a^{d/2}}w\Big(\theta^{-1}\Big(\frac{\bm{x}'-\bm{x}}{a}\Big)\Big),
    \quad
    \Tilde{w}_{a,\theta}(\bm{p}) 
    = a^{d/2}\Tilde{w}(\theta^{-1}(a\bm{p})),
    \quad
    \Tilde{w}(\bm{p}) 
    = \int \dd[d]{\bm{x}}\ee^{\ii\bm{p}\cdot\bm{x}}w(\bm{x}).
    \label{eq:continuous_wavelet_transform_functions}
\end{equation}
The wavelet function must be square integrable and must satisfy the admissibility condition
\begin{equation}
    C_w = \int_{\mathbb{R}^d}\frac{\abs{\Tilde{w}(\bm{p})}^2}{\abs{\bm{p}}^d}\dd[d]{\bm{p}}<+\infty,
    \label{eq:admission}
\end{equation}
which then ensures that the wavelet transform can be inverted using the reconstruction formula
\begin{equation}
    \phi(\bm{x}) 
    = \frac{1}{C_w} \int_0^{+\infty} \int_{SO(d)} \int_{\mathbb{R}^d}
    w_{a,\theta,\bm{x}'}(\bm{x}) \phi(\bm{x}',a,\theta)
    \frac{\dd[d]{\bm{x}'} \dd{\mu(\theta)} \dd{a}}{a^{d+1}}.
\end{equation}
In this paper we work primarily with isotropic wavelet functions that are independent of orientation $\theta$.
This is because most objects that we will transform--quantum field correlators--will also depend on the spatial distance and not on the orientation of the spatial vector.
Subsequently, this will allow us to integrate out the angular part of the Fourier transform, resulting in a Hankel transform.

Now, following from the definition \cref{eq:continuous_wavelet_transform_manyd}, we will be interested in the behaviour of wavelet-transformed integral kernels $K(\bm{x},\bm{y},a,b)$
\begin{align}
    \nonumber K(\bm{x},\bm{y},a,b) 
    &= \int_{\mathbb{R}^{2d}}w_a^*(\bm{x}'-\bm{x})K(\bm{x}',\bm{y}')w_b(\bm{y}'-\bm{y})\dd[d]{\bm{x}'}\dd[d]{\bm{y}'} \\
    &= \int_{\mathbb{R}^{2d}}\ee^{-\ii\bm{p}\cdot\bm{x}}\Tilde{w}_a^*(\bm{p})\Tilde{K}(\bm{p},\bm{q})\Tilde{w}_b(\bm{q})\ee^{\ii\bm{q}\cdot\bm{y}}\frac{\dd[d]{\bm{p}}\dd[d]{\bm{q}}}{(2\pi)^{2d}}.
\end{align}
When the kernel is a convolution kernel $K(\bm{x},\bm{y}) = K(\bm{x}-\bm{y})$, the expression simplifies to
\begin{equation}
    K(\bm{x}-\bm{y},a,b) 
    = \int_{\mathbb{R}^{d}}\ee^{-\ii\bm{p}\cdot(\bm{x}-\bm{y})}\Tilde{w}_a^*(\bm{p})\Tilde{K}(\bm{p})\Tilde{w}_b(\bm{p})\frac{\dd[d]{\bm{p}}}{(2\pi)^{d}}.
\end{equation}
For \cref{eq:bosonic_wavelet_correlator_manyd} we need expansions of isotropic kernels $\Tilde{K}(\bm{p}) = \Tilde{K}(\abs{\bm{p}})$ around their diagonal $\bm{x}=\bm{y}$. 
This can be obtained via the Hankel transform:
\begin{align}
    \nonumber K(\bm{x}-\bm{y},a,b) 
    &= \frac{1}{(2\pi)^\frac{d}{2}\abs{\bm{x}-\bm{y}}^{\frac{d}{2}-1}}\mathcal{H}_{\frac{d}{2}-1}\big[p^{\frac{d}{2}-1}\Tilde{w}^*_{a}(p)\Tilde{K}(p)\Tilde{w}_b(p)\big](\abs{\bm{x}-\bm{y}})\\
    &= \sum_{k=0}^{+\infty}\frac{(-1)^k\Gamma(\frac{d}{2})}{k!\,\Gamma(\frac{d}{2}+k)}\left(\frac{1}{\pi^\frac{d}{2}\Gamma(\frac{d}{2})}\int_0^{+\infty}p^{2k}\Big(\frac{p}{2}\Big)^{d-1}\Tilde{w}^*_{a}(p)\Tilde{K}(p)\Tilde{w}_b(p)\dd{p}\right)\Big(\frac{\abs{\bm{x}-\bm{y}}}{2}\Big)^{2k},
    \label{eq:WaveletHankel}
\end{align}
where $\mathcal{H}_\nu[f(p)](r) = \int_0^{+\infty}f(p)J_\nu(pr)p\dd{p}$ is defined for $\nu>1/2$ and $J_\nu$ is the Bessel function of the first kind.
The expansion formula remains valid for $d = 1$ when the wavelet is either an odd or even function.
In our QFT applications we will often deal with kernels of the following form
\begin{equation}
    \Tilde{K}(p) = 
    (p^2+m^2)^{l/2}
    \left\{\begin{array}{cc}
        1 & l\text{ even} \\
        f_\beta\big(\sqrt{p^2+m^2}\big) & l\text{ odd}
    \end{array}\right.,
    \quad
    f_\beta(\omega) =
    \left\{\begin{array}{cc}
        \coth\frac{\beta\omega}{2} & \text{for bosons} \\
        \tanh\frac{\beta\omega}{2} & \text{for fermions}
    \end{array}\right..
\end{equation}
For convenience, we will introduce the following coefficients
\begin{equation}
    M_{k,l}^{(i,j)}(a,b;\beta;m) =  
    \frac{a^{-i}b^{-j}}{\pi^{\frac{d}{2}}\Gamma(\frac{d}{2})}\int_0^{+\infty}\dd{p}p^{k}(p^2+m^2)^{l/2}\left(\frac{p}{2}\right)^{d-1}(\,(\ii\hat{D})^{\underline{i}}\,\Tilde{w})^*_a(p)(\,(\ii\hat{D})^{\underline{j}}\,\Tilde{w})_b(p)
    \times\left\{\begin{array}{cc}
        1 & l\text{ even} \\
        f_\beta\big(\sqrt{p^2+m^2}\big) & l\text{ odd}
    \end{array}\right.,
    \label{eq:WaveletMDefinition_Appendix}
\end{equation}
where $\hat{D}=\frac12(\hat{\bm{p}}\cdot\hat{\bm{x}}+\hat{\bm{x}}\cdot\hat{\bm{p}})$ is the dilation operator, and $\hat{x}^{\underline{i}}=\hat{x}(\hat{x}-I)\cdots(\hat{x}-(i+1)I)$ is the falling factorial. The indices $i$ and $j$ correspond to derivatives with respect to scales
\begin{equation}
    \partial_a^i\partial_b^j M_{k,l}^{(0,0)}(a,b;\beta;m) = M_{k,l}^{(i,j)}(a,b;\beta;m),
\end{equation}
which explains the appearance of $(\ii\hat{D})^{\underline{i}}$, because
\begin{equation}
    \partial_a^i \Tilde{w}_a(p) = \partial_a^i \ee^{\ii\log a \hat{D}}\Tilde{w}(p) = a^{-i}\ee^{\ii\log a \hat{D}}(\ii\hat{D})^{\underline{i}}\,\Tilde{w}(p) = a^{-i}((\ii\hat{D})^{\underline{i}}\,\Tilde{w})_a(p).
\end{equation}
These coefficients satisfy
\begin{align}
    M_{k,l}^{(i,j)}(a,b;\beta;m) &= a^{-(i+j+k+l)} M_{k,l}^{(i,j)}(1,b/a;\beta/a;am) = b^{-(i+j+k+l)} M_{k,l}^{(i,j)}(a/b,1;\beta/b;bm),
    \label{eq:wavelet_M_ab}
    \\
    M_{k,l}^{(i,j)}(a,a;\beta;m) &= a^{-(i+j+k+l)} M_{k,l}^{(i,j)}(1,1;\beta/a;am),
    \label{eq:wavelet_M_aa}
    \\
    M_{k,l}^{(i,j)}(a,b;\beta;0) &= M_{k+l,0}^{(i,j)}(a,b;\beta;0).
\end{align}
Some of our results can be expressed in terms of quantities dependent only on the wavelet
\begin{equation}
    W_k^{(i)}(a,b) 
    = \matrixel{w_a}{\abs{\hat{\bm{p}}}^k(\ii\hat{D})^i}{w_b}
    = \frac{1}{\pi^{\frac{d}{2}}\Gamma(\frac{d}{2})}\int_0^{+\infty}p^{k}\left(\frac{p}{2}\right)^{d-1}\Tilde{w}^*_a(p)((\ii\hat{D})^i\Tilde{w})_b(p)\dd{p}.
    \label{eq:wavelet_onesided_momentum_moment_app}
\end{equation}
These are the limiting values of the coefficients defined above
\begin{align}
    M_{k,l}^{(0,0)}(a,b;\infty;0) &= W_{k+l}^{(0)}(a,b)
    \\
    M_{k,l}^{(i,j)}(a,b;\infty;0) &= a^{-i}b^{-j}\sum_{p=0}^i\sum_{q=0}^j \frac{1}{p!}\dv[p]{\lambda}(-\lambda^{\underline{i}})\Big|_{\lambda=k+l} s(j,q) W_{k+l}^{(p+q)}(a,b),
\end{align}
where $s(n,m)$ are the Stirling numbers of the first kind
\begin{equation}
    \lambda^{\underline{n}}=\sum_{m=0}^n s(n,m)\lambda^m.
\end{equation}
It can be easily checked that they similarly satisfy (while also introducing shorthand notation $W_k^{(i)} \equiv W_k^{(i)}(1,1)$):
\begin{align}
    W_k^{(i)}(a,b) &= a^{-k} W_k^{(i)}(1,b/a) = b^{-k} W_k^{(i)}(a/b,1),
    \label{eq:wavelet_onesided_momentum_moment_ab}
    \\
    W_k^{(i)}(a,a) &= a^{-k} W_k^{(i)}(1,1) \equiv a^{-k} W_k^{(i)},
    \label{eq:wavelet_onesided_momentum_moment_aa}
    \\
    W_0^{(0)}(a,a)
    &= W_0^{(0)}(1,1)
    \equiv W_0^{(0)}
    = 1.
    \label{eq:wavelet_onesided_momentum_moment_zero}
\end{align}

%

\section{Scale-dependent QFT}
\label{app:wavelet_correlators}

\subsection{Free boson}
\label{app:wavelet_correlators_bosonic}

The action of a Hermitian scalar field theory in $D \equiv 1+d$ dimensional spacetime is
\begin{equation}
    S[\phi] 
    = \int \dd[D]{x} \left[ -\frac12 \eta^{\mu\nu}\partial_\mu \phi(x) \partial_\nu \phi(x) - \frac12 m^2 \phi(x)^2 \right],
\end{equation}
with $\eta_{\mu\nu}=\operatorname{diag}(-1,1,\ldots,1)$. The Euler-Lagrange equation is then
\begin{equation}
    (-\eta^{\mu\nu}\partial_\mu \partial_\nu + m^2) \phi(x) = 0,
\end{equation}
and in flat spacetime, the field and the conjugate field operators have the mode expansions
\begin{align}
    \hat{\phi}(x) &= \int \frac{\dd[d]{\bm{p}}}{(2\pi)^{d}} \frac{1}{2 \omega_{\bm{p}}}(\hat{a}(\bm{p}) \ee^{\ii p_\mu x^\mu} + \hat{a}^\dag(\bm{p}) \ee^{-\ii p_\mu x^\mu}),
    \\
    \hat{\pi}(x) &= \int \frac{\dd[d]{\bm{p}}}{(2\pi)^{d}} \frac{-\ii}{2}(\hat{a} (\bm{p}) \ee^{\ii p_\mu x^\mu} - \hat{a}^\dag(\bm{p}) \ee^{-\ii p_\mu x^\mu}),
\end{align}
where $p^0 = -p_0 = \omega_{\bm{p}} = \sqrt{\bm{p}^2+m^2}$. 
We then have the following equal-time commutation relations
\begin{align}
    [\hat{\phi}(t,\bm{x}), \hat{\pi}(t,\bm{y})]
    &= \ii \delta(\bm{x} - \bm{y}),
    \\
    [\hat{a}(\bm{p}), \hat{a}^\dag(\bm{p}')]
    &= (2\pi)^d 2\omega_{\bm{p}} \delta(\bm{p} - \bm{p}').
\end{align}
Using the thermal expectation values of the creation and annihilation operators (see \cite{salvio2025})
\begin{align}
    \expval*{\hat{a}(\bm{p})\hat{a}^\dag(\bm{p}')}_\beta &= (2\pi)^d2\omega_{\bm{p}}(1+n_\beta(\omega_{\bm{p}}))\delta(\bm{p}-\bm{p}'),\\
    \expval*{\hat{a}^\dag(\bm{p})\hat{a}(\bm{p}')}_\beta &= (2\pi)^d2\omega_{\bm{p}}n_\beta(\omega_{\bm{p}})\delta(\bm{p}-\bm{p}'),\\
    \expval*{\hat{a}(\bm{p})\hat{a}(\bm{p}')}_\beta &= \expval*{\hat{a}^\dag(\bm{p})\hat{a}^\dag(\bm{p}')}_\beta = 0,
\end{align}
the thermal correlation function is then given by
\begin{equation}
    \expval*{\hat{\phi}(x)\hat{\phi}(y)}_\beta = \int\frac{\dd[d]{\bm{p}}}{(2\pi)^d}\frac{1}{2\omega_{\bm{p}}}[(1+n_\beta(\omega_{\bm{p}}))\ee^{\ii p_\mu(x^\mu-y^\mu)}+n_\beta(\omega_{\bm{p}})\ee^{-\ii p_\mu(x^\mu-y^\mu)}],
\end{equation}
where $n_\beta(\omega)=\frac{1}{\ee^{\beta\omega}-1}$ is the thermal occupation factor which in the zero temperature case ($\beta=\infty$) simplifies to the usual vacuum 2-point correlation function
\begin{equation}
    \expval{\hat{\phi}(x) \hat{\phi}(y)}
    = \int \frac{\dd[d]{\bm{p}}}{(2\pi)^d}
        \frac{\ee^{\ii p_\mu (x^\mu - y^\mu)}}{2\omega_{\bm{p}}}
    = \begin{cases}
        -\dfrac{1}{4\pi} \log \abs{s^2}
        & m=0,\;d=1,
        \\
        \dfrac{\Gamma(\frac{d-1}{2})}{4\pi^{\frac{d+1}{2}}}\abs{s^2}^{\frac{1-d}{2}}
        & m=0,\;d>1,
        \\
        \dfrac{1}{(2\pi)^\frac{d+1}{2}}\left(\dfrac{m^2}{\abs{s^2}}\right)^\frac{d-1}{4}K_\frac{d-1}{2}(m\sqrt{\abs{s^2}}) 
        & m \neq 0,
    \end{cases}
\end{equation}
where $s^2 = \eta_{\mu\nu}(x^\mu-y^\mu)(x^\nu-y^\nu)$. 
Applying the continuous wavelet transform to the field and its conjugate in the spatial dimensions results in
\begin{align}
    \hat{\phi}(t,a,\theta,\bm{x}) 
    &= \int \dd[d]{\bm{y}} w_{a,\theta}(\bm{y}-\bm{x})^*\,\hat{\phi}(t,\bm{y})
    = \int \frac{\dd[d]{\bm{p}}}{(2\pi)^{d}}\frac{1}{2\omega_{\bm{p}}}(\hat{a}(\bm{p})\Tilde{w}_{a,\theta}(\bm{p})^*\ee^{\ii p_\mu x^\mu}+\hat{a}^\dag(\bm{p})\Tilde{w}_{a,\theta}(\bm{p})\ee^{-\ii p_\mu x^\mu}),
    \\
    \hat{\pi}(t,a,\theta,\bm{x}) 
    &= \int \dd[d]{\bm{y}} w_{a,\theta}(\bm{y}-\bm{x})^*\,\hat{\pi}(t,\bm{y})
    = \int \frac{\dd[d]{\bm{p}}}{(2\pi)^{d}}\frac{-\ii}{2}(\hat{a}(\bm{p})\Tilde{w}_{a,\theta}(\bm{p})^*\ee^{\ii p_\mu x^\mu}-\hat{a}^\dag(\bm{p})\Tilde{w}_{a,\theta}(\bm{p})\ee^{-\ii p_\mu x^\mu}),
\end{align}
where $a>0$ is the scale and $\theta \in SO(d)$ is the orientation as per \cref{eq:continuous_wavelet_transform_manyd}. The two-point functions in the wavelet picture read
\begin{align}
    \nonumber\expval*{\hat{\phi}(x,a,\theta_1)\hat{\phi}(y,b,\theta_2)}_\beta 
    &= \int \dd[d]{\bm{x}'}\dd[d]{\bm{y}'} w_{a,\theta_1}(\bm{x}'-\bm{x})\langle\hat{\phi}(x^0,\bm{x}')\hat{\phi}(y^0,\bm{y}')\rangle w_{b,\theta_2}(\bm{y}'-\bm{y})
    \\
    &= \int \frac{\dd[d]{\bm{p}}}{(2\pi)^d}\ee^{-\ii\bm{p}\cdot(\bm{x}-\bm{y})}\Tilde{w}_{a,\theta_1}(\bm{p})^*\frac{1}{2\omega_{\bm{p}}}\Big[(1+n_\beta(\omega_{\bm{p}}))\ee^{-\ii \omega_{\bm{p}}(x^0-y^0)}+n_\beta(\omega_{\bm{p}})\ee^{\ii \omega_{\bm{p}}(x^0-y^0)}\Big]\Tilde{w}_{b,\theta_2}(\bm{p}),
    \\
    \nonumber\expval*{\hat{\pi}(x,a,\theta_1)\hat{\pi}(y,b,\theta_2)}_\beta
    &= \int \dd[d]{\bm{x}'}\dd[d]{\bm{y}'} w_{a,\theta_1}(\bm{x}'-\bm{x})\langle\hat{\pi}(x^0,\bm{x}')\hat{\pi}(y^0,\bm{y}')\rangle w_{b,\theta_2}(\bm{y}'-\bm{y})
    \\
    &= \int \frac{\dd[d]{\bm{p}}}{(2\pi)^d}\ee^{-\ii\bm{p}\cdot(\bm{x}-\bm{y})}\Tilde{w}_{a,\theta_1}(\bm{p})^*\frac{\omega_{\bm{p}}}{2}\Big[(1+n_\beta(\omega_{\bm{p}}))\ee^{-\ii \omega_{\bm{p}}(x^0-y^0)}+n_\beta(\omega_{\bm{p}})\ee^{\ii \omega_{\bm{p}}(x^0-y^0)}\Big]\Tilde{w}_{b,\theta_2}(\bm{p}).
\end{align}
For isotropic wavelets we can expand the anti-commutator functions around the coincidence $y\to x$ as
\begin{align}
    \expval*{\{\hat{\phi}(x,a),\hat{\phi}(y,b)\}}_\beta
    &= \frac12\sum_{m=0}^{+\infty}\sum_{n=0}^{+\infty}\frac{(-1)^{m+n}\Gamma(\frac{d}{2})}{(2m)!n!\,\Gamma(\frac{d}{2}+n)}M_{2n,2m-1}^{(0,0)}(a,b;\beta;m)(x^0-y^0)^{2m}\big(\tfrac12|\bm{x}-\bm{y}|\big)^{2n},\\
    \expval*{\{\hat{\pi}(x,a),\hat{\pi}(y,b)\}}_\beta
    &= \frac12\sum_{m=0}^{+\infty}\sum_{n=0}^{+\infty}\frac{(-1)^{m+n}\Gamma(\frac{d}{2})}{(2m)!n!\,\Gamma(\frac{d}{2}+n)}M_{2n,2m+1}^{(0,0)}(a,b;\beta;m)(x^0-y^0)^{2m}\big(\tfrac12 |\bm{x}-\bm{y}|\big)^{2n},
\end{align}
with $M_{k,l}^{(i,j)}(a,b;\beta;m)$ defined in \cref{eq:WaveletMDefinition_Appendix} with $f_\beta(\omega) = \coth\tfrac12\beta\omega$ accounting for the bosonic statistics.

\subsection{Free fermion}
\label{app:wavelet_correlators_fermionic}

The action of a free Dirac fermion in $D \equiv 1+d$ dimensional spacetime is
\begin{equation}
    S[\bar{\Psi},\Psi]
    = \int \dd[D]{x} \bar{\Psi}(x)(\gamma^\mu \partial_\mu - m) \Psi(x),
\end{equation}
where $\gamma^\mu$ are the Dirac $\gamma$-matrices satisfying
\begin{equation}
    \frac12\{\gamma^\mu,\gamma^\nu\} 
    = \eta^{\mu\nu}\mathbb{1},\quad\eta^{\mu\nu}=\operatorname{diag}(-1,1,\ldots,1).
    \label{eq:gamma_algebra}
\end{equation}
Variation with respect to the adjoint $\bar{\Psi} \equiv \Psi^\dagger \ii\gamma^0$ gives the Dirac equation
\begin{equation}
    (\gamma^\mu \partial_\mu - m) \Psi(x) 
    = 0.
\end{equation}
Following the same idea as in the case of the scalar field one can obtain the thermal correlation matrix \cite{salvio2025}
\begin{equation}
    \expval*{\hat{\Psi}(x) \hat{\bar{\Psi}}(y)}_\beta
    = \int \frac{\dd[d]{\bm{p}}}{(2\pi)^d}
    \frac{1}{2\omega_{\bm{p}}}\left[(1-n_\beta(\omega_{\bm{p}}))\Big( \ii p_\mu\gamma^\mu + m \Big)
    \ee^{\ii p_\mu(x^\mu-y^\mu)}+n_\beta(\omega_{\bm{p}})\Big( \ii p_\mu\gamma^\mu - m \Big)\ee^{-\ii p_\mu(x^\mu-y^\mu)}\right],
\end{equation}
where $p_0=-\omega_{\bm{p}}$, and $n_\beta(\omega)=\frac{1}{\ee^{\beta\omega}+1}$ is the thermal occupation factor.
After the wavelet transform, we obtain the correlation matrix
\begin{align}
    \nonumber&\expval*{\hat{\Psi}(x,a,\theta_1) \hat{\bar{\Psi}}(y,b,\theta_2)}_\beta
    = \\
    &\int \frac{\dd[d]{\bm{p}}}{(2\pi)^d}
    \Tilde{w}^*_{a,\theta_1}(\bm{p})\frac{1}{2\omega_{\bm{p}}}\left[(1-n_\beta(\omega_{\bm{p}}))\Big( \ii p_\mu\gamma^\mu + m \Big)
    \ee^{\ii p_\mu(x^\mu-y^\mu)}+n_\beta(\omega_{\bm{p}})\Big( \ii p_\mu\gamma^\mu - m \Big)\ee^{-\ii p_\mu(x^\mu-y^\mu)}\right]\Tilde{w}_{b,\theta_2}(\bm{p}).
\end{align}
Since the correlation matrix of the Dirac field is not a Lorentz scalar it could potentially be interesting to consider non-isotropic wavelets. 
However, for simplicity and consistency we will continue to restrict ourselves to isotropic wavelets.

For a massless theory in $d=1$, the Clifford algebra \cref{eq:gamma_algebra} is satisfied by the $\gamma$-matrices
\begin{equation}
    \gamma^0 = \begin{pmatrix}
        0 & -1 \\
        1 &  0
    \end{pmatrix}, \quad
    \gamma^1 = \begin{pmatrix}
        0 & 1 \\
        1 & 0
    \end{pmatrix},
\end{equation}
and we are mainly interested in the correlator $\expval*{\hat{\Psi}(t,x,a) \hat{\Psi}^\dagger(t',y,b)}_\beta$, which can be obtained straightforwardly via
\begin{align}
    &\nonumber\expval*{\hat{\Psi}(t,x,a) \hat{\Psi}^\dagger(t',y,b)}_\beta
    = \expval*{\hat{\Psi}(t,x,a) \hat{\bar{\Psi}}(t',y,b)}_\beta (- \ii \gamma^0)
    =\int\frac{\dd{p}}{2\pi}\ee^{\ii p(x-y)}\Tilde{w}_a(p)^*\Tilde{w}_b(p)
    \\
    &\begin{pmatrix}
        (1-n_\beta(\omega_p))\frac12(1+\frac{p}{\omega_p})\ee^{-\ii\omega_p (t-t')}+n_\beta(\omega_p)\frac12(1-\frac{p}{\omega_p})\ee^{\ii\omega_p(t-t')} & \frac{\ii m}{2\omega_p}((1-n_\beta(\omega_p))\ee^{-\ii\omega_p(t-t')}-n_\beta(\omega_p)\ee^{\ii\omega_p(t-t')}) \\
        -\frac{\ii m}{2\omega_p}((1-n_\beta(\omega_p))\ee^{-\ii\omega_p(t-t')}-n_\beta(\omega_p)\ee^{\ii\omega_p(t-t')}) & (1-n_\beta(\omega_p))\frac12(1-\frac{p}{\omega_p})\ee^{-\ii\omega_p (t-t')}+n_\beta(\omega_p)\frac12(1+\frac{p}{\omega_p})\ee^{\ii\omega_p(t-t')}
    \end{pmatrix}
\end{align}
The expansion of the correlation function around the coincidence limit $(t',y)\to(t,x)$ to second order is then
\begin{align}
    \nonumber\expval*{\hat{\Psi}(t,x,a) \hat{\Psi}^\dagger(t',y,b)}_\beta&=\frac{1}{2}\Big(M_{0,0}^{(0,0)}(a,b;\beta;m)-im M_{0,-1}^{(0,0)}(a,b;\beta;m)\gamma^0
    \\
    \nonumber&\quad\quad-\ii (M_{0,1}^{(0,0)}(a,b;\beta;m) + m M_{0,0}^{(0,0)}(a,b;\beta;m)\gamma^0)t+\ii M_{2,-1}^{(0,0)}(a,b;\beta;m)\gamma^1\gamma^0x
    \\
    \nonumber&\quad\quad\quad-\frac{1}{2}(M_{0,2}^{(0,0)}(a,b;\beta;m)-\ii m M_{0,1}^{(0,0)}(a,b;\beta;m)\gamma^0)t^2-\frac12 (M_{2,0}^{(0,0)}(a,b;\beta;m)-\ii m M_{2,-1}^{(0,0)}(a,b;\beta;m)\gamma^0)x^2
    \\
    &\quad\quad\quad\quad+M_{2,0}^{(0,0)}(a,b;\beta;m)\gamma^1\gamma^0tx\Big)
\end{align}
where $M_{k,l}^{(i,j)}(a,b;\beta;m)$ is defined in \cref{eq:WaveletMDefinition_Appendix} with $f_\beta(\omega) = \tanh\tfrac12\beta\omega$.

%

\section{Metric reconstruction formula for the Dirac field}
\label{app:metric_from_dirac}

Here we present the derivation of a corresponding formula to \cref{eq:metric_from_scalar_field_correlations}, in the case of the massless Dirac field on Minkowski spacetime.
We begin with the correlation function for the massless scalar field, which is
\begin{equation}
    \expval{\hat{\phi}(x)\hat{\phi}(y)}
    = \dfrac{\Gamma(\frac{d-1}{2})}{4\pi^{\frac{d+1}{2}}}\abs{s^2}^{\frac{1-d}{2}}.
\end{equation}
The correlation matrix for the Dirac field can be obtained by applying the Dirac differential operator to the scalar field correlation function
\begin{align}
    \nonumber\expval*{\hat{\Psi}(x) \hat{\bar{\Psi}}(y)}
    &= \gamma^\mu\partial_\mu \expval{\hat{\phi}(x) \hat{\phi}(y)}
    \\
    \nonumber&= \dfrac{\Gamma(\frac{d-1}{2})}{4\pi^{\frac{d+1}{2}}} \gamma^\mu \partial_\mu \abs{s^2}^{\frac{1-d}{2}}
    \\
    \nonumber&= \frac{ \frac{1-d}{2}\Gamma(\frac{d-1}{2}) }{ 4\pi^{\frac{d+1}{2}} } \abs{s^2}^{\frac{-1-d}{2}}
    2\eta_{\mu\nu}\gamma^\mu(x^\nu-y^\nu)
    \\
    &=-\frac{\Gamma(\frac{d+1}{2})}{2\pi^{\frac{d+1}{2}}} 
    \frac{\eta_{\mu\nu}\gamma^\mu(x^\nu-y^\nu)}{\abs{s^2}^\frac{d+1}{2}},
\end{align}
which also holds for the case $d=1$. 
Using the properties of the gamma matrices it follows that
\begin{equation}
    \expval*{\hat{\Psi}(x) \hat{\bar{\Psi}}(y)}^2
    = \frac{\Gamma(\frac{d+1}{2})^2}{4\pi^{d+1}} \frac{s^2}{\abs{s^2}^{d+1}} \mathbb{1}.
    \label{eq:dirac_correlations_squared}
\end{equation}
Thus we can recover Synge's world function, which is
\begin{equation}
    \frac{1}{2}s^2 
    = \frac{1}{2} \left( \frac{\Gamma(\frac{d+1}{2})}{2\pi^{\frac{d+1}{2}}}\right)^{\frac{2}{d}}
    \left[\frac{ \Tr\left(\expval*{\hat{\Psi}(x)\hat{\bar{\Psi}}(y)}^2\right) }{ \Tr\left(\mathbb{1}\right) }\right]^{-\frac{1}{d}}.
\end{equation}
Note that the matrix trace is not the only possible basis-invariant operation for inverting the identity matrix, one could also consider the determinant.
However, whereas the determinant is related to the modulus of a Clifford algebra element in a highly nontrivial manner, the trace has a straightforward interpretation: it can be interpreted as a scalar product of Clifford algebra elements, and thus generalises easily to higher dimensions.
Finally, the metric is then
\begin{equation}
    g_{\mu\nu}(x)
    = -\frac{1}{2}\left(\frac{\Gamma(\frac{d+1}{2})}{2\pi^{\frac{d+1}{2}}}\right)^{\frac{2}{d}}
    \lim_{y\to x} \partial_\mu \partial'_\nu \left[\frac{\Tr\left(\expval*{\hat{\Psi}(x)\hat{\bar{\Psi}}(y)}^2\right)}{\Tr\left(\mathbb{1}\right)}\right]^{-\frac{1}{d}},
    \label{eq:metric_from_dirac_field_correlations_app}
\end{equation}
and a similar argument as made by \textcite{saravani2016} then follows as to why mass and curvature will not impact the leading order behaviour of $\expval*{\hat{\Psi}(x) \hat{\bar{\Psi}}(y)}^2$ around coincidence so that \cref{eq:metric_from_dirac_field_correlations_app} may generalise to any curved manifold.

%

\section{Petz-Rényi mutual information}
\label{app:prmi}

In the main text, the $\alpha=2$ Petz-Rényi mutal information was used to determine the induced metric. For Gaussian states, the PRMI can be derived from the state's covariance matrices, affording certain advantages if $\alpha=2$, as noted below.

\subsection{Bosonic Gaussian state}
\label{app:prmi_bosonic}

Following \textcite{casini2018}, the Petz-Rényi mutual information for subsystems of a bosonic Gaussian state is
\begin{equation}
    I_\alpha = \frac{1}{2(1-\alpha)}\log\left(\frac{\det(T_\text{j}^\alpha-T_\text{p}^{\alpha-1})}{\det(T_\text{j}-1)^\alpha\det(T_\text{p}-1)^{1-\alpha}}\right),
    \label{eq:prmi_bosonic_app}
\end{equation}
where
\begin{equation}
    T_\text{j}=\begin{pmatrix}
        1 & 0 \\
        0 & P_\text{j}
    \end{pmatrix}\begin{pmatrix}
        \frac{C_\text{j}^2+\frac14}{C_\text{j}^2-\frac14} & \ii \frac{C_\text{j}^2}{C_\text{j}^2-\frac14} \\
        -\ii \frac{1}{C_\text{j}^2-\frac14} & \frac{C_\text{j}^2+\frac14}{C_\text{j}^2-\frac14}
    \end{pmatrix}\begin{pmatrix}
        1 & 0 \\
        0 & P_\text{j}^{-1}
    \end{pmatrix},
    \quad
    T_\text{p}=\begin{pmatrix}
        1 & 0 \\
        0 & P_\text{p}
    \end{pmatrix}\begin{pmatrix}
        \frac{C_\text{p}^2+\frac14}{C_\text{p}^2-\frac14} & \ii \frac{C_\text{p}^2}{C_\text{p}^2-\frac14} \\
        -\ii \frac{1}{C_\text{p}^2-\frac14} & \frac{C_\text{p}^2+\frac14}{C_\text{p}^2-\frac14}
    \end{pmatrix}\begin{pmatrix}
        1 & 0 \\
        0 & P_\text{p}^{-1}
    \end{pmatrix}.
\end{equation}
Here $C_\text{j}$ is the correlation matrix of the joint state
\begin{equation}
     C_{\text{j}}^2
     = X_{\text{j}}P_{\text{j}},
     \quad
     X_{\text{j}}
     = \frac12 \begin{pmatrix}
        \expval*{\{\hat{\phi}_A,\hat{\phi}_A\}}_\beta & \expval*{\{\hat{\phi}_A,\hat{\phi}_B\}}_\beta
        \\ 
        \expval*{\{\hat{\phi}_A,\hat{\phi}_B\}}_\beta & \expval*{\{\hat{\phi}_B,\hat{\phi}_B\}}_\beta
    \end{pmatrix},
    \quad
    P_{\text{j}}
    = \frac12 \begin{pmatrix}
        \expval*{\{\hat{\pi}_A,\hat{\pi}_A\}}_\beta & \expval*{\{\hat{\pi}_A,\hat{\pi}_B\}}_\beta
        \\ 
        \expval*{\{\hat{\pi}_A,\hat{\pi}_B\}}_\beta & \expval*{\{\hat{\pi}_B,\hat{\pi}_B\}}_\beta
    \end{pmatrix},
 \end{equation}
 and $C_\text{p}$ is the correlation matrix for the product state
 \begin{equation}
     C_{\text{p}}^2
     = X_{\text{p}} P_{\text{p}},
     \quad
     X_{\text{p}}
     = \begin{pmatrix}
        \expval*{\hat{\phi}_A\hat{\phi}_A}_\beta & 0
        \\ 
        0 & \expval*{\hat{\phi}_B\hat{\phi}_B}_\beta
        \end{pmatrix},\quad
    P_{\text{p}}
    = \begin{pmatrix}
        \expval*{\hat{\pi}_A\hat{\pi}_A}_\beta & 0
        \\ 
        0 & \expval*{\hat{\pi}_B\hat{\pi}_B}_\beta
        \end{pmatrix}.
 \end{equation}
The meaningful range for the parameter is $\alpha \in [0,2]$ (with $\alpha=1$ defined by the limit) because in this range the Petz-Rényi relative entropy is monotonic under quantum channels \cite{PETZ198657}. The non-integer matrix powers in \cref{eq:prmi_bosonic_app} are defined in terms of spectral decomposition, that is, diagonalising the matrices and then raising the eigenvalues to that power. However, this is only well-defined when the eigenvalues of the matrix are positive. Denoting $t$ the eigenvalues of the matrix $T$ and $\sigma^2$ the eigenvalues of $C^2$ we can see that they are related by
\begin{equation}
    t_\pm = \frac{\sigma^2+\frac14}{\sigma^2-\frac14}\pm\sqrt{\left(\frac{\sigma^2+\frac14}{\sigma^2-\frac14}\right)^2-1} = \frac{(\sigma\pm\sgn(\sigma-\frac12)\frac12)^2}{(\sigma+\frac12)(\sigma-\frac12)}.
\end{equation}
The implicit assumption $\sigma^2\geq0$ ensures that the eigenvalues are real. We can see that the eigenvalues $t$ are negative when $\sigma<\frac12$ and positive when $\sigma>\frac12$. This condition is often called the physicality condition.

An issue arises when we introduce the wavelet representation. In the coincidence limit, the wavelets will inevitably have overlap, which  violates the physicality condition for the joint covariance matrix.
This leads us to the choice $\alpha=2$ for the Petz-Rényi mutual information to avoid raising negative numbers to non-integer powers.

Substituting $\alpha=2$ into \cref{eq:prmi_bosonic_app}, we have
\begin{align}
    \nonumber I_2
    &= -\frac{1}{2} \log \left( \frac{\det(T_{\text{j}}^2 - T_{\text{p}})}{\det(T_{\text{j}} - 1)^2\det(T_{\text{p}} - 1)^{-1}} \right)
    \\
    &= -\frac{1}{2} \log \left( \frac{1+8(X_{1,2}P_{1,2} - \frac12(X_{1,1} P_{1,1} + X_{2,2} P_{2,2})) + 16 ((X_{1,2})^{2} - X_{1,1} X_{2,2})((P_{1,2})^{2} - P_{1,1} P_{2,2})}{(1 - 4 X_{1,1} P_{1,1})^2 (1 - 4 X_{2,2} P_{2,2})^2} \right),
\end{align}
where $X_{i,j} = [X_{\text{j}}]_{ij}$ and $P_{i,j}=[P_{\text{j}}]_{ij}$.

\subsection{Fermionic Gaussian state}
\label{app:prmi_fermionic}

As per \textcite{casini2018}, the PRMI for a Gaussian state of a fermionic QFT is expressible as
\begin{equation}
    I_\alpha = -\log\det(1-C_{\text{p}}) - \frac{\alpha}{1-\alpha}\log\det(1-C_{\text{j}}) - \frac{1}{1-\alpha}\log\det(1 + \Big(\frac{C_{\text{j}}}{1-C_{\text{j}}}\Big)^\alpha\Big(\frac{C_{\text{p}}}{1-C_{\text{p}}}\Big)^{1-\alpha}),
\end{equation}
where $C_{\text{j}}$ and $C_{\text{p}}$ can be written in a block form as
\begin{equation}
    C_{\text{j}} 
    = \begin{pmatrix}
        \expval*{\hat{\Psi}_A\hat{\Psi}_{A}^\dag} & \expval*{\hat{\Psi}_A\hat{\Psi}_B^\dag} \\
        \expval*{\hat{\Psi}_B\hat{\Psi}_{A}^\dag} & \expval*{\hat{\Psi}_B\hat{\Psi}_{B}^\dag}
    \end{pmatrix},\quad
    C_{\text{p}} 
    = \begin{pmatrix}
        \expval*{\hat{\Psi}_A\hat{\Psi}^\dag_A} & 0 \\
        0 & \expval*{\hat{\Psi}_B\hat{\Psi}^\dag_B}
    \end{pmatrix}.
\end{equation}
For $\alpha=2$ we obtain 
\begin{equation}
    I_2(A;B)
    = \log(\frac{\det(C_\text{j}^2-2C_\text{j}C_\text{p}+C_\text{p})}{\det(C_\text{p}(1-C_\text{p}))}).
\end{equation}

\section{Affine group coherent state wavelets}
\label{app:affine_group_wavelets}

Here we introduce a family of wavelets that will provide a simple parameterization for the emergent metric derived from field correlations derived above.
We consider the $(1+1)$-dimensional case but wavelets of this form for higher dimensions can also be constructed. 

The affine group $\mathrm{Aff}(\mathbb{R})\cong \mathbb{R}\rtimes \mathbb{R}^+$ is the set of transformations $x\rightarrow a x+b$ with $a>0$, b$\in\mathbb{R}$.
It has two generators, the dilation operator $\hat{D}=\frac{1}{2}(\hat{x}\hat{p}+\hat{p}\hat{x})$ which generates rescaling by a positive number, and the momentum operator $\hat{p}$ which generates shifts.
In order to construct coherent states we seek a canonically conjugate pair of operators.
Under dilations
\begin{equation}
    \ee^{-\ii \lambda \hat{D}}\hat{p} \ee^{\ii \lambda \hat{D}}
    = \ee^{\lambda}\hat{p},\quad \lambda \in \mathbb{R},
\end{equation}
which suggests that if we want an operator that is shifted by a real number when acted on by dilations it should be of logarithmic form. 
The operator $\hat{u}=\log(\abs{\hat{p}})$ suffices since
\begin{align}
    \nonumber \ee^{-\ii \lambda \hat{D}} \hat{u} \ee^{\ii \lambda \hat{D}}
    &= \log(\abs{\ee^{\lambda}\hat{p}})
    \\
    \nonumber&= \log(\ee^{\lambda}\abs{\hat{p}})
    \\
    &= \lambda + \hat{u}.
\end{align}
Indeed one can write $\hat{D}=\ii \frac{\partial}{\partial u}$
\footnote{
    A direct translation of co\"ordinates would appear to produce $\hat{D} = \ii(\frac{\partial}{\partial u}+\frac{1}{2})$ which is not Hermitian.
    However, there is a change of measure when integrating wavefunctions.
    Namely, for $\psi(p) \in L^2(\mathbb{R})$ the momentum representation of $\phi(u) \in L^2(\mathbb{R})$ we have $\ee^{u/2}\psi(\ee^u)=\phi(u)$, so the representation $\hat{D}=\ii\frac{\partial}{\partial u}$ is the correct one.
}, and we have the canonical commutation relation: $[\hat{D},\hat{u}]=\ii$.
The coherent states with respect to this canonical pair are of Gaussian form
\begin{equation}
    \phi(u)
    = \frac{1}{\sqrt{\pi^{1/2}\sigma}} \ee^{-\frac{1}{2\sigma^2}(u-u_0)^2-\ii D_0 (u-u_0)},
\end{equation}
where $u_0=\langle \hat{u}\rangle$ and $D_0=\langle \hat{D}\rangle$ and $\sigma\in(0,+\infty)$ where the boundary values do not limit to proper coherent states.
One easily finds that
\begin{equation}
    \langle \hat{u}^2 \rangle 
    = u_0^2+\sigma^2/2,
    \quad
    \langle {D}^2 \rangle 
    = D_0^2 + \frac{1}{2\sigma^2}, 
\end{equation}
confirming that these are indeed minimum uncertainty states satisfying $\Delta(\hat{D}) \Delta(\hat{u}) = \frac12$.
We can rewrite such a coherent state as a square integrable function over the momentum variable for $p \geq 0$, by making the replacement $u=\log p$ and noting the measure is $\dd{p}=\ee^u \dd{u}$,
\begin{equation}
    \psi(p)
    = \frac{1}{\sqrt{\pi^{1/2}\sigma}}\frac{1}{\sqrt{p}}\ee^{-\frac{1}{2\sigma^2}(\log{p}-u_0)^2 - \ii D_0 (\log{p}-u_0)}, 
    \quad p\geq 0
\end{equation}
This can be extended to a function over all values of $p$, where we additionally take $u_0=0=D_0$, as follows
\begin{equation}
    \psi(p)
    = \frac{\sgn{p}}{\sqrt{2\pi^{1/2}\sigma}}\frac{1}{\sqrt{\abs{p}}}\ee^{-\frac{1}{2\sigma^2}(\log{\abs{p}})^2}.
\end{equation}
This function will allow us to construct a wavelet in momentum space if we introduce an additional scale degree of freedom $a>0$ such that
\begin{equation}
    \psi_a(p) = \sqrt{a}\psi\Big(pa\Big).
\end{equation}
Making the substitution, the following function achieves this:
\begin{equation}
    \tilde{w}_a(p)
    = \frac{\sgn{p}}{\sqrt{2\pi^{1/2}\sigma}}\frac{1}{\sqrt{\abs{p}}}\ee^{-\frac{1}{2\sigma^2}(\log{\abs{p}}+\log{a})^2}.
\end{equation}
We call this an affine group coherent state wavelet.
It is an antisymmetric function in $p$ and satisfies the admissibility condition with $C_w=\ee^{\sigma^2/4}$ as well as $\int_{\mathbb{R}}\abs{\tilde{w}_a(p)}^2 \dd{p}
    = 1$.

\begin{figure}
    \centering
    \includegraphics[width=.8\columnwidth]{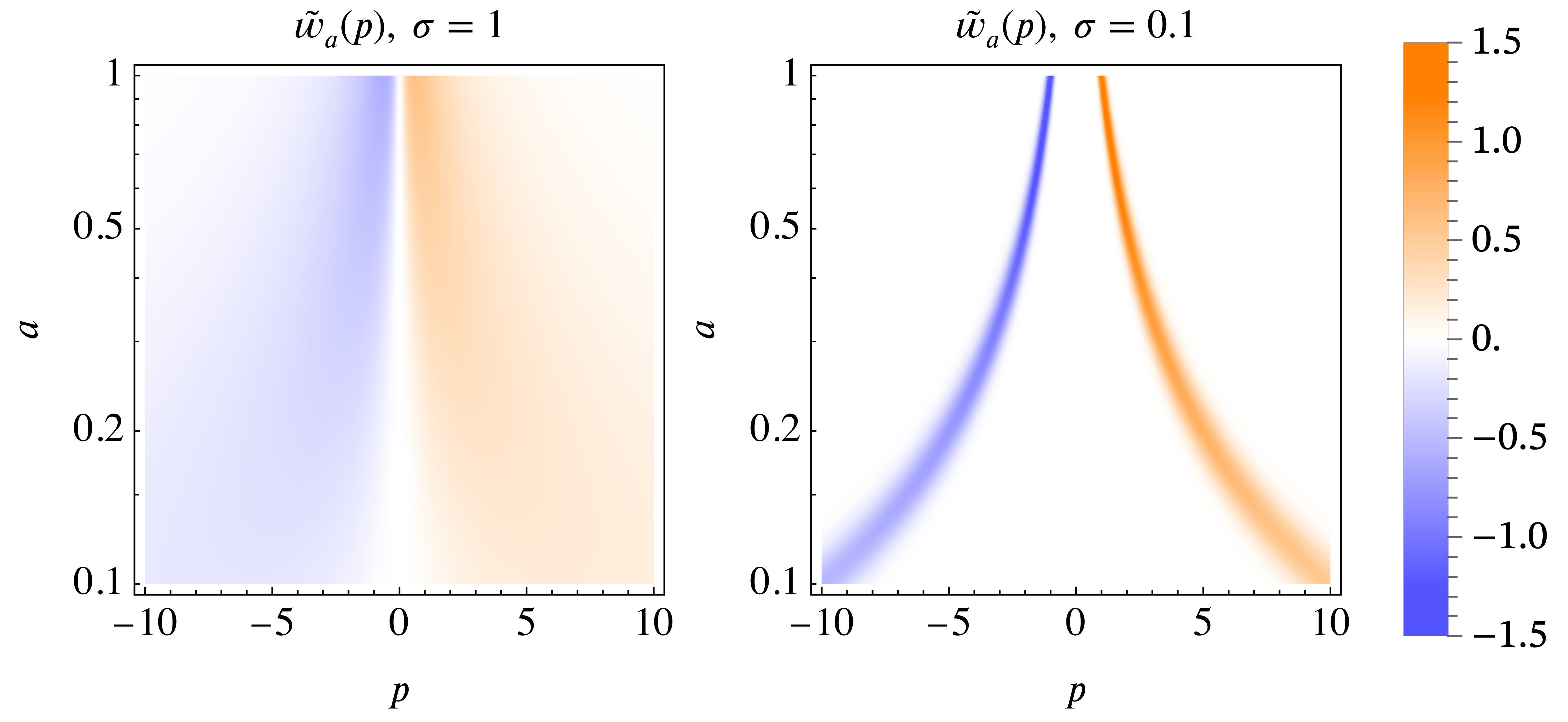}
    \caption{Density plots of the affine group coherent state wavelets as a function of momentum and scale.
    This wavelet family is parameterised by $\sigma$, the standard deviation of the co\"ordinate $u=\log(\abs{\hat{p}})$ conjugate to the dilation operator $\hat{D}$.
    Plot of $\tilde{w}_a(p)$ for $\sigma=1$ (left), $\sigma=0.1$ (right).}
\label{fig:AffineWavelet}
\end{figure}

For computations of moments of the dilation and momentum operators, it is convenient to express the wavelet in the $u$ co\"ordinate. In this case, we lose information about $\sgn(p)$ so this definition is appropriate for $p>0$:
\begin{equation}
    w_a(u) 
    = \frac{1}{\sqrt{2\pi^{1/2}\sigma}}\ee^{-(u+\log{a})^2/(2\sigma^2)}.
\end{equation}
Note $w_a(u)$ is sub-normalized so that $\int_{\mathbb{R}} \abs{w_a(u)}^2 \dd{u} = \frac{1}{2}$. Then we can compute
\begin{align}
    \nonumber
    W^{(\ell)}_k(a,a)
    &= \matrixel{w_a}{\abs{\hat{p}}^k(\ii \hat{D})^\ell}{w_{a}}
    \\
    \nonumber
    &= 2\int_{0}^{\infty}\tilde{w}^*_a(p) p^k (\ii \hat{D})^\ell \tilde{w}_a(p) \dd{p}
    \\
    &= (-1)^\ell 2\int_{-\infty}^{\infty}w^*_a(u) \ee^{ku} (\frac{\partial}{\partial u})^\ell w_a(u) \dd{u}.
\end{align}
Changing variables to $v=u+\log(a)$,
\begin{align}
    \nonumber
    W^{(\ell)}_k(a,a)
    &= \frac{(-1)^la^{-k}}{\sqrt{\pi} \sigma}\int_{-\infty}^{\infty}
    \ee^{-\frac{v^2}{2\sigma^2}} \ee^{kv} \Big(\frac{\partial}{\partial v}\Big)^\ell \ee^{-\frac{v^2}{2\sigma^2}} \dd{v}
    \\
    \nonumber
    &= \frac{a^{-k}}{\sqrt{\pi} \sigma }\int_{-\infty}^{\infty} \frac{1}{(\sqrt{2} \sigma)^\ell} H_\ell\left( \frac{v}{\sqrt{2}\sigma} \right) \ee^{kv} \ee^{-\frac{v^2}{\sigma^2}} \dd{v}
    \\
    &= \frac{a^{-k}\sqrt{2}}{\sqrt{\pi} (\sqrt{2} \sigma)^\ell}\int_{-\infty}^{\infty} H_\ell(z) \ee^{-2z^2+k\sqrt{2}\sigma z} \dd{z},
\end{align}
where $H_\ell$ are the Hermite polynomials and in the last line we changed variables again to $z=v/\sqrt{2}\sigma$.

Now, consider the generating function for the Hermite polynomials, $\sum_{\ell=0}^{\infty} H_\ell(z)\frac{t^\ell}{\ell!} = \ee^{-t^2+2t z}$.
Weighting both sides by the exponential $\ee^{-2z^2+k\sqrt{2} \sigma z}$ and integrating over $z$ gives:
\begin{align}
    \sum_{\ell=0}^{\infty}\frac{t^\ell}{\ell!}\int_{-\infty}^{\infty}H_\ell(z) \ee^{-2z^2+k\sqrt{2}\sigma z} \dd{z}
    &= \sqrt{\frac{\pi}{2}} \ee^{\frac{1}{8}(k\sqrt{2}\sigma)^2 + \frac12 (k\sqrt{2}\sigma t - t^2)}
    \\
    &= \sqrt{\frac{\pi}{2}} \ee^{\frac{1}{4} k^2 \sigma^2} \sum_{m=0}^{\infty} H_m \left( \frac{k\sigma}{2} \right) \frac{t^m}{2^{m/2}m!}.
\end{align}
Comparing like powers of $t$, we find that
\begin{equation}
    \int_{-\infty}^{\infty} H_\ell(z) \ee^{-2z^2+k\sqrt{2}\sigma z} \dd{z}
    = \sqrt{\frac{\pi}{ 2^{\ell+1}}} \ee^{\frac{k^2\sigma^2}{4}} H_\ell\left( \frac{k\sigma}{2} \right).
\end{equation}
And so:
\begin{equation}
    W^{(\ell)}_k(a,a)
    =\frac{a^{-k}}{2^\ell \sigma^\ell} \ee^{\frac{1}{4} k^2 \sigma^2} H_\ell \left( \frac{k\sigma}{2} \right).
\end{equation}

\end{document}